\documentclass[numberedappendix,appendixfloats]{emulateapj}
\usepackage{amsmath,graphicx}

\shortauthors{T. Hosokawa et al.}
\shorttitle{Evolution of Massive Protostars via Disk Accretion}

\begin{document}

\title{Evolution of Massive Protostars via Disk Accretion}

\author{Takashi Hosokawa\altaffilmark{1,2,3},  
        Harold W. Yorke\altaffilmark{2},
        Kazuyuki Omukai\altaffilmark{1,3}}

\altaffiltext{1}{Department of Physics, Kyoto University, 
Kyoto 606-8502, Japan; hosokawa@tap.scphys.kyoto-u.ac.jp, omukai@tap.scphys.kyoto-u.ac.jp}
\altaffiltext{2}{Jet Propulsion Laboratory, California Institute
of Technology, Pasadena CA 91109; Takashi.Hosokawa@jpl.nasa.gov, 
Harold.Yorke@jpl.nasa.gov}
\altaffiltext{3}{Division of Theoretical Astronomy, 
National Astronomical Observatory, Mitaka, Tokyo 181-8588, Japan}

\begin{abstract}
Mass accretion onto (proto-)stars at high accretion rates 
$\dot{M}_\ast > 10^{-4}~M_{\odot}~{\rm yr}^{-1}$ is expected
in massive star formation. We study the evolution of massive protostars at 
such high rates by numerically solving the stellar structure equations.
In this paper we examine the evolution via disk accretion.
We consider a limiting case of ``cold'' disk accretion, 
whereby most of the stellar photosphere 
can radiate freely with negligible backwarming from the accretion flow,
and the accreting material settles onto the
star with the same specific entropy as the photosphere.
We compare our results to the calculated evolution via 
spherically symmetric accretion, the opposite limit, whereby the 
material accreting onto the star contains the
entropy produced in the accretion shock front. 
We examine how different accretion geometries 
affect the evolution of massive protostars.
For cold disk accretion at $10^{-3}~M_{\odot}~{\rm yr}^{-1}$ 
the radius of a protostar is initially small, $R_\ast \simeq$ a few $R_\odot$. 
After several solar masses have accreted, the protostar begins to bloat up 
and for $M_\ast \simeq 10~M_\odot$ the stellar radius attains its 
maximum of $30 - 400~R_\odot$. 
The large radius $\sim$$100~R_\odot$ is also a feature of
spherically symmetric accretion at the same accreted mass and accretion rate. 
Hence, expansion to a large radius is a robust feature of accreting
massive protostars. At later times the protostar eventually begins
to contract and reaches the Zero-Age
Main-Sequence (ZAMS) for $M_\ast \simeq 30~M_\odot$, 
independent of the accretion geometry. 
For accretion rates exceeding several
$10^{-3}~M_{\odot}~{\rm yr}^{-1}$
the protostar never contracts to the ZAMS.
The very large radius of several $100{\rm s}~R_\odot$ results in
a low effective temperature and low UV luminosity of the protostar.
Such bloated protostars could well explain the existence of bright high-mass
protostellar objects, which lack detectable H~II regions.
\end{abstract}

\keywords{accretion -- stars: early-type -- stars: evolution 
-- stars: formation -- stars: pre-main sequence}

\section{Introduction}
\label{sec:intro}

Recent studies have revealed that the formation process of 
massive $(> 8~M_\odot)$ stars differ in many respects from that of low-mass 
$(\sim 1~M_\odot)$ stars 
(e.g., Zinnecker \& Yorke 2007).
Although signatures of mass accretion are widely observed toward 
forming massive stars, the estimated accretion rates usually exceed 
$10^{-4}~M_{\odot}~{\rm yr}^{-1}$, 
which are much higher than the typical value in low-mass star 
formation,
\begin{equation}
\dot{M}_\ast \sim \frac{c_s^3}{G} \simeq 
5 \times 10^{-6} \left( \frac{T}{10~{\rm K}} \right)~ 
M_\odot~{\rm yr}^{-1},
\label{eq:lmass_acc}
\end{equation}
where $c_s$ and $T$ is the sound speed and temperature in 
natal molecular cloud cores, and $G$ is the gravity constant.
In the past it has been argued that 
such high accretion rates are required for accretion flow
to overcome the strong radiation pressure from forming massive stars
(e.g., Larson \& Starrfield 1971; Wolfire \& Cassinelli 1987).
However, these results were based on the assumption of spherically
symmetric infall.  Nevertheless, there are other good reasons to
expect high accretion rates during the formation of massive stars:
Massive stars have short Kelvin-Helmholz time scales, once formed
they consume the available nuclear fuel quickly, and they will lose
a significant part of their mass through strong stellar winds.  Thus, 
massive stars evolve quickly, even while accreting.  We can expect the 
time spent in the main accretion phase to be significantly less than 
the main sequence lifetime, presumably of the order of a few dynamical
times of the star-forming molecular core.  For a $30~M_\odot$ star
this provides a lower limit to the average accretion rate
$\simeq 10^{-4}~M_{\odot}~{\rm yr}^{-1}$.  
Accretion rates may vary strongly, however.
Thus, phases of even higher accretion rates are likely.

The high accretion rates suggest that the nature of
massive star formation differs from that of low-mass star formation.
So far, various formation scenarios, such as monolithic collapse
of massive dense cores 
(e.g., Yorke \& Sonnhalter 2002; McKee \& Tan 2002, 2003; 
Krumholz et al. 2007, 2009), and the competitive accretion with 
global infall of the cluster-forming clumps 
(e.g., Bonnell et al. 1998; Bonnell, Vine \& Bate 2004; Wang et al. 2010), 
have been proposed. 
The proposed formation scenarios are being assessed 
with high-resolution observations of the massive cores and clumps 
(e.g., Motte et al. 2007; Marseille et al. 2008; Zhang et al. 2009; 
Bentemps et al. 2010).

The evolution of massive protostars at such high rates 
is a critical feature of massive star formation which determines
the effect these stars have on their environment
(e.g., Zinnecker \& Yorke 2007).
In our previous work (Hosokawa \& Omukai 2008; 2009, 
hereafter Paper I), we examined the evolution by numerically
solving the interior structure of a protostar with a dusty 
accretion envelope under the assumption of spherical symmetry.  
Our calculations show that accreting massive protostars have 
certain characteristic features. 
At $\dot{M}_\ast = 10^{-3}~M_{\odot}~{\rm yr}^{-1}$, for example,
the protostellar radius becomes very large, exceeding $100~R_\odot$
at maximum. The evolution before the arrival to the zero-age main 
sequence (ZAMS) lasts until $M_\ast \simeq 30~M_\odot$.
At even higher accretion rates 
$\dot{M}_\ast > 3 \times 10^{-3}~M_{\odot}~{\rm yr}^{-1}$, 
the protostar very abruptly inflates and overtakes material in the
inner part the accreting envelope, 
thus invalidating the assumption of
steady mass accretion before reaching the ZAMS.

The very large stellar radius leads to a low effective 
temperature of the protostar, which in turn explains some
observational properties of massive protostars in Orion KL nebula 
(Morino et al. 1998; Furuya \& Shinnaga 2009).

Observations suggest that mass 
accretion onto massive stars is not spherically symmetric, but rather
proceeds through massive circumstellar 
disks (e.g., Cesaroni et al. 2007).
Signatures of rotating infall are widely found
toward forming massive stars, sometimes associated with massive outflows
along the rotation axis (e.g., Patel et al. 2005; Beltr\'an et al. 2006a).
Recent numerical simulations also suggest that the massive disk is a
natural outcome of collapse of a massive dense core
(Yorke \& Sonnhalter 2002; Krumholz et al. 2007, 2009).
Protostellar evolution via disk accretion has been studied
for lower-mass protostars with low accretion rates 
$\dot{M}_\ast \leq 10^{-4}~M_{\odot}~{\rm yr}^{-1}$
(e.g., Palla \& Stahler 1992; Beech \& Mitalas 1994; Hartman et al. 1997).
These authors considered the limiting case of ``cold'' disk accretion,
whereby most of the stellar surface is able to radiate into free space
and the accreting material brings only a small amount of entropy into the star.
They showed that, with the low accretion rates, 
disk accretion slightly reduces the stellar radius
compared to the spherically symmetric cases.

Maeder and coworkers studied the evolution of accreting massive
protostars in the cold disk limit with mass accretion rates
increasing with the stellar mass up to $10^{-4}~M_\odot~{\rm yr}^{-1}$
(e.g., Norberg \& Maeder 2000; Behrend \& Maeder 2001).
Recently, Yorke \& Bodenheimer (2008, hereafter YB08) calculated
the evolution in the same limit with much higher accretion rates
$10^{-3}$ and $10^{-2}~M_\odot~{\rm yr}^{-1}$.
Their calculated evolution differs from the spherical accretion
case in an early phase of $M_\ast \lesssim 8~M_\odot$. 
The stellar radius is several 10s$~R_\odot$ with
spherical accretion, whereas it is a few $R_\odot$ with disk accretion.
However, YB08 find that the stellar radius rapidly increases
soon afterwards and exceeds $100~R_\odot$ at $M_\ast \simeq 10~M_\odot$.
The subsequent evolution at $M_\ast \gtrsim 10~R_\odot$
is quite similar for both spherical accretion and disk accretion.

In this paper, we examine how the different accretion 
geometries affect the evolution of massive protostars at 
high accretion rates. We explain why the two extreme cases
lead to the same characteristic features of massive protostars, 
such as the large radius, in spite of the different evolution in 
the early phases. 
To this end, we calculate the protostellar evolution
in the cold disk accretion limit with the numerical code 
used in Paper I.
This allows us to extract how the protostellar evolution depends
on the different accretion geometry independent of the effects of
different stellar evolution codes.  For completeness, we also compare 
the calculated evolution with previous results by YB08.

Finally, we stress that this study is very relevant to star formation
in the early universe, where high accretion rates are also expected.
Protostellar evolution with high accretion rates has been 
studied with zero or very low metallicities under the assumption of
spherically symmetric accretion
(e.g., Stahler, Palla \& Salpeter 1986; Omukai \& Palla 2001, 2003; 
Hosokawa \& Omukai 2009b). 
Recent numerical simulations show that circumstellar accretion disks 
are also expected during star formation in the early universe 
(Yoshida, Omukai \& Hernquist 2008; Stacy, Grief \& Bromm 2010).
Tan \& McKee (2004) examined the evolution of primordial protostars
via disk accretion using simple analytic models.
Here, we calculate the detailed evolution of a primordial 
protostar in the cold disk accretion limit.

The organization of this paper is as follows. 
The numerical method and the cases considered are briefly
described in Section \ref{sec:num}.
In Section \ref{sec:result}, we focus on the protostellar
evolution at the accretion rate 
$\dot{M}_\ast = 10^{-3}~M_{\odot}~{\rm yr}^{-1}$, which allows a
direct comparison to the results of YB08.
First, we briefly review the evolution via spherical accretion studied
in Paper I (Section \ref{ssec:sp_acc}), 
and then consider the evolution via cold disk accretion 
(Section \ref{ssec:d_acc}). 
Evolution of a primordial protostar is investigated in Section 
\ref{ssec:metal} and subsequently contrasted to the present-day cases.
Variation of the protostellar evolution with different accretion rates
is studied in Section \ref{sec:mdot_dep}.
Finally, Section \ref{sec:sum} is devoted to our summary and conclusions. 

\begin{figure*}
\plotone{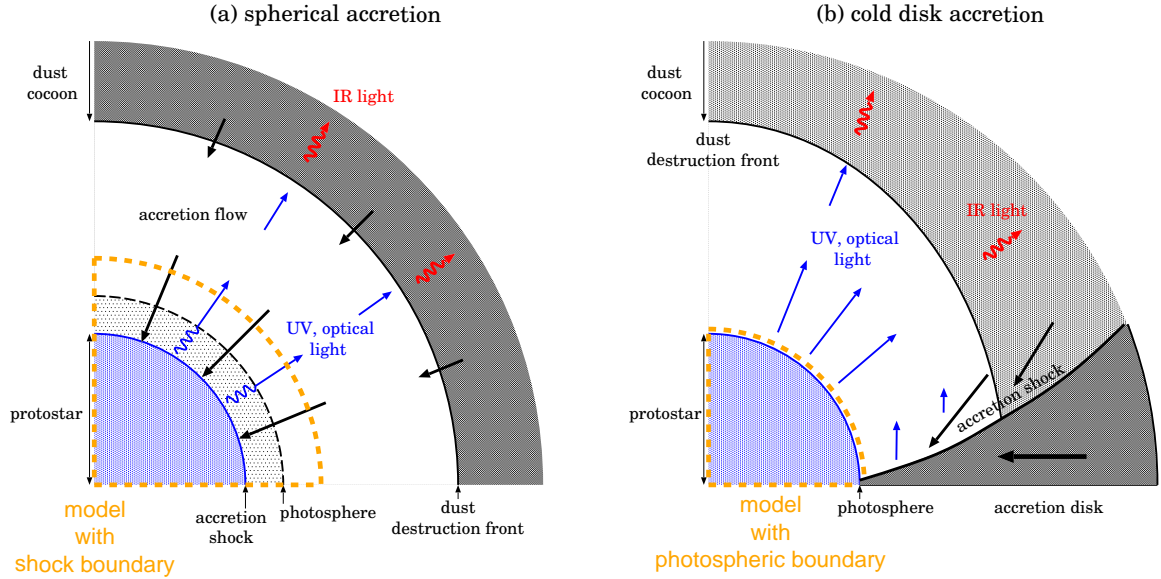}
\caption{Schematic figures of a protostar with different
accretion geometry: (a) spherically symmetric accretion, 
and (b) cold disk accretion.
In the spherical case (a), the accretion flow directly hits the
stellar surface forming an accretion shock front. If the flow is
optically thick before reaching the surface, the photosphere 
locates outside the stellar surface. 
Dust grains in the accretion envelope evaporate in a dust
destruction front far outside the photosphere.
In the cold disk accretion case (b),
gas predominantly accretes onto a circumstellar disk rather than the star.
Accreting material settles onto the stellar surface through a
geometrically thin layer (or possibly through geometrically thin
accretion columns -- not shown).  Heat brought into the star
in the accretion flow radiates freely into space until the temperature
attains the photospheric value.
Most of the stellar surface is unaffected by the accretion flow. The
energy radiated away by the disk and/or accretion columns before
the material settles onto the star (the so-called ``accretion luminosity'') 
needs to be accounted for separately from the intrinsic stellar
luminosity.}
\label{fig:prost_schem_spph}
\end{figure*}

\section{Numerical Modeling}
\label{sec:num}

\subsection{Modeling Mass Accretion via Disks}
\label{ssec:coldisk}

We examine protostellar evolution by numerically
solving the stellar structure equations. 
Case (a) in Figure \ref{fig:prost_schem_spph} presents
the schematic picture of the spherical accretion studied in 
Paper I, where we assumed spherical symmetry for an
accretion envelope as well as the protostar.
The numerical method we adopted is summarized as follows 
(e.g., Stahler, Shu \& Taam 1980; also see Appendix A of Paper I).
For a protostar, we solve the four stellar structure equations 
taking into account mass accretion. 
Free-fall flow is assumed for the optically thin part of the
accretion envelope.
If the accretion flow becomes opaque to the stellar radiation
before hitting the stellar surface, we solve for the 
structure of this optically thick part of the flow together with
the stellar structure equations of the protostar and the
jump conditions at the accretion shock front. 
The accretion shock front is the outer boundary 
of the modeled protostar (shock boundary). 
 
For the case of mass accretion 
via a circumstellar disk we consider the limiting case whereby the disk-star 
connecting layer is geometrically thin and most of the stellar 
surface is unlinked to the accretion flow.
Figure \ref{fig:prost_schem_spph} depicts 
this situation.
In this case, the usual photospheric outer boundary condition
can be applied for accreting protostars 
(e.g., Palla \& Stahler 1992), 
\begin{equation}
\label{eq:psb_p}
P_{\rm sf} = \frac23 \frac{1}{\kappa_{\rm sf}} \frac{G M_\ast}{R_\ast^2} ,
\end{equation}
\begin{equation}
\label{eq:psb_l}
L_\ast = 4 \pi R_\ast^2 \sigma T_{\rm sf}^4 ,
\end{equation}
where the suffix "sf" indicates the values at the stellar surface, 
and $\sigma$ is Stefan-Boltzman constant.
In this limiting case of disk accretion, accreting materials add
to the star with the smallest amount of entropy, which is the
value in the stellar atmosphere unattached to the accretion flow.
Hence, the treatment with equations (\ref{eq:psb_p}) and
(\ref{eq:psb_l}) supposes the limit of ``cold'' disk accretion.
Note that this cold disk accretion is just a limiting case.
Gas accreting onto the star through the disk is likely to have
somewhat higher entropy especially at high accretion rates 
(Hartmann et al. 1997). 
The spherical accretion is the opposite limit where accreting 
materials carry the highest amount entropy into the star. 
Thus, we consider two extreme limiting cases with the 
spherical accretion and cold disk accretion. 
In order to solve the protostellar structure with cold disk accretion,
we modified the numerical code used in Paper I to deal with
the photospheric boundary condition.
We do not solve the detailed structure of the flow connecting 
the star and disk.

Different geometries of the accretion flow will lead to different
structures of the evolving protostar. 
For example, the average entropy in the stellar interior
reflects the history of entropy brought into the star with 
accreting material, which in turn depend on the
accretion geometry. For spherical accretion at high accretion
rates the accretion shock front is embedded inside the 
stellar photosphere and most of the
high entropy created at the accretion shock front 
is carried to the stellar 
surface. This entropy is efficiently taken into the 
stellar interior with high accretion rates (Paper I).
As a result the protostar has a higher average entropy
(compared to a non-accreting protostar), which leads to a very large
radius -- exceeding $100~R_\odot$ at maximum.
With disk accretion, however, entropy is mainly 
generated within the disk by viscous dissipation. 
Before the accreted material reaches the stellar surface, a large fraction 
of the generated entropy will be transported away by radiation.
As a result, entropy brought into the star with the accreting material
is significantly lower than in the spherical accretion case.
The average entropy within the star $s$ is related to 
the stellar radius as 
\begin{equation}
R_{\ast} \propto M_{\ast}^{-1/3}{\rm exp}[const. \times s].
\label{eq:r_entropy}
\end{equation}

This relation is derived by substituting the typical gas density 
and pressure within a star of mass $M_{\ast}$ and radius $R_{\ast}$ 
(e.g., Cox \& Giuli 1968);
\begin{equation}
\rho \sim \frac{M_{\ast}}{R_{\ast}^3}, 
\qquad P \sim G \frac{M_{\ast}^2}{R_{\ast}^4},
\label{eq:typ_rhop}
\end{equation}
to the definition of specific entropy of ideal monotonic gas,
\begin{equation}
s = \frac{3 \cal{R}}{2 \mu} \ln \left( \frac{P}{\rho^{5/3}} \right) + const.,
\label{eq:s_gene}
\end{equation}
where ${\cal R}$ is the gas constant and $\mu$ is the mean
molecular weight.
Equation (\ref{eq:r_entropy}) demonstrates that, for the same stellar mass, 
the stellar radius is larger with the higher average entropy 
in the stellar interior. 
Thus, we can naively expect that the stellar radius is 
reduced with disk accretion. 
Contrary to this expectation, however, calculations by 
YB08, adopting the cold disk accretion, 
show that the radius of a massive protostar
exceeds $100~R_\odot$ at maximum as with the spherical accretion.
The goal of this paper is to explain why the outcomes are so 
similar among the extreme cases of the spherical accretion and
cold disk accretion.

\begin{table*}[t]
\label{tb:md}
\begin{center}
Table 1. Input Parameters of the cases considered \\[3mm]
\begin{tabular}{l|ccccccc}
\hline
Case    & $\dot{M}_\ast~(M_{\odot}/{\rm yr})^a$ & geometry 
       & $Z^c$ & $M_{\ast,0}~(M_{\odot})^d$ & $R_{\ast,0}~(R_{\odot})^e$ 
       & reference$^f$ \\
\hline
\hline
MD4x3-SDm0.5$^g$  & $4 \times 10^{-3}$  &  S $\to$ D  & 0.02  & 0.3 (0.5) 
              & 32.7 (36.8) & Section \ref{sec:mdot_dep}  \\ 
MD4x3-S       & $4 \times 10^{-3}$  &  S  & 0.02  & 0.3  & 32.7
                       & Section \ref{sec:mdot_dep}, Paper I  \\ 

MD3-D      & $10^{-3}$  &  D  & 0.02  & 0.1  & 3.7
                       & Section~\ref{ssec:d_acc}  \\ 
MD3-D-b0.1 & $10^{-3}$  &  D  & 0.02   & 0.1 & 3.5
                       & Section~\ref{sssec:imod} \\
MD3-D-cv  & $10^{-3}$  &  D  & 0.02  &  1.3  & 6.2
                       & Section~\ref{sssec:imod}  \\
MD3-S     & $10^{-3}$  & S   & 0.02   & 0.05 & 15.5
                       & Section~\ref{ssec:sp_acc}, Paper I \\
MD3-S-z0  & $10^{-3}$  & S   & 0.0  &  0.05  & 13.0
                       & Section~\ref{ssec:metal}, Paper I \\
MD3-SDm1  & $10^{-3}$  & S $\to$ D   & 0.02   & 0.05 (1.0) & 15.5 (24.4)
                       &  Section~\ref{sssec:imod}  \\
MD3-SDm0.1  & $10^{-3}$  & S $\to$ D   & 0.02   & 0.05 (0.1) & 15.5 (14.1)
                       & Section~\ref{sssec:imod}  \\
MD3-SDm0.1-z0  & $10^{-3}$  & S $\to$ D   & 0.0   & 0.05 (0.1) & 13.0 (14.1)
                       & Section~\ref{ssec:metal} \\
MD4-S       & $10^{-4}$  & S & 0.02   & 0.03  & 8.8 
                         & Section~\ref{sec:mdot_dep}, Paper I \\
MD4-SDm0.1  & $10^{-4}$  & S $\to$ D   & 0.02   & 0.03 (0.1) & 8.8 (5.1)
                       & Section~\ref{sec:mdot_dep} \\
MD4-D-cv   & $10^{-4}$  & D   & 0.02  &  1.3  & 6.2
                       & Appendix~\ref{ap:cp_ps} \\
MD5-D-cv   & $10^{-5}$  & D   & 0.02  &  1.0  & 4.3
                       & Appendix~\ref{ap:cp_ps} \\
MD5-S-cv  & $10^{-5}$  & S   & 0.02  &  1.0  & 4.2
                         &  Appendix~\ref{ap:cp_ps}, Paper I \\
\hline                           
\end{tabular}
\noindent
\end{center} 
$a$ : mass accretion rate, 
$b$ : geometry of the accretion flow
(D: disk accretion, S: spherical accretion), 
$c$ : metallicity, $d$ : mass of initial core model,
$e$ : radius of initial core model,  
$f$ : subsections where numerical results of each case are presented. 
$g$ : the suffix SDm$X$ means the geometry is switched 
      from spherical accretion to disk accretion at $M_\ast = X~M_\odot$.
      For these cases, stellar mass and radius at the switching point
      are listed in the brackets in the columns of $M_{\ast,0}$ and $R_{\ast,0}$.
\end{table*}

\subsection{Cases considered}

Table 1 summarizes the cases considered in this study and their 
input parameters. 
For simplicity the protostellar evolution is calculated 
with a constant given accretion rate for each case.  
The adopted accretion rates range from $10^{-5}~M_\odot~{\rm yr}^{-1}$ 
up to $4 \times 10^{-3}~M_\odot~{\rm yr}^{-1}$.
Evolution for the accretion rate $\dot{M}_\ast = 10^{-3}~M_{\odot}~{\rm yr}^{-1}$
is studied in detail.
In addition to evolution with cold disk accretion we present several
cases with spherical accretion for comparison.
Evolution via disk accretion at the rates 
$\dot{M}_\ast = 10^{-5}~M_{\odot}~{\rm yr}^{-1}$ and
$10^{-4}~M_{\odot}~{\rm yr}^{-1}$ was also studied by 
Palla \& Stahler (1992). 
We consider protostellar evolution with these low rates to
compare our results with theirs (see Appendix~\ref{ap:cp_ps}).

The initial model in each case is constructed following 
Stahler, Shu \& Taam (1980) or Palla \& Stahler (1991)
with the adopted accretion rate and boundary conditions 
(also see Appendix A.2 and B.2 in Paper I).
The initial stellar mass $M_{\ast,0}$ is taken as an arbitrary small value. 
The initial entropy profile in the stellar interior is assumed 
as a function of the mass coordinate $M$, 
\begin{equation}
s(M) = s_{c,0} 
+ \beta \frac{k_{\rm B}}{m_{\rm H}} \frac{M}{M_{\ast,0}} ,
\label{eq:s0}
\end{equation}
where $s_{c,0}$ is specific entropy at the stellar center, 
$k_{\rm B}$ is Boltzmann constant, $m_{\rm H}$ is atomic mass 
unit, and $\beta~(> 0)$ is a free parameter. 
We adopt $\beta = 1$ as a fiducial value. 
Unlike the cases of spherical accretion, however, we will see that 
the protostellar evolution with the photospheric boundary
changes with different initial models even at the same 
accretion rate (see Section \ref{sssec:imod} below). 
We show this by presenting the evolution with different 
initial models with $\beta = 0.1$ 
(shallower entropy profile, case MD3-D-b0.1), 
and $\beta = 0$ (homogeneous entropy profile, cases with the suffix ``cv''). 

The geometry of the accretion flow will vary as a function of the stellar
mass and radius. Initially, the collapse of a gravitationally unstable
molecular core is mostly spherically symmetric.  
A hydrostatic object is formed when the central regions of 
the molecular core become optically thick.
A circumstellar disk is later formed as material with increasingly higher 
angular momentum falls toward the newly formed protostar.
For these cases we start the protostellar evolution
calculation assuming spherical accretion and later switch
to cold disk accretion (cases with the suffix ``SD'').  
We also present the evolution of primordial protostars 
via disk accretion (case MD3-SDm0.1-z0).

\section{Evolution of Massive Protostars with
$\dot{M}_\ast = 10^{-3}~M_{\odot}~{\rm yr}^{-1}$}
\label{sec:result}

\subsection{Spherically Symmetric Accretion}
\label{ssec:sp_acc}

\begin{figure*}[t]
\epsscale{0.7}
\plotone{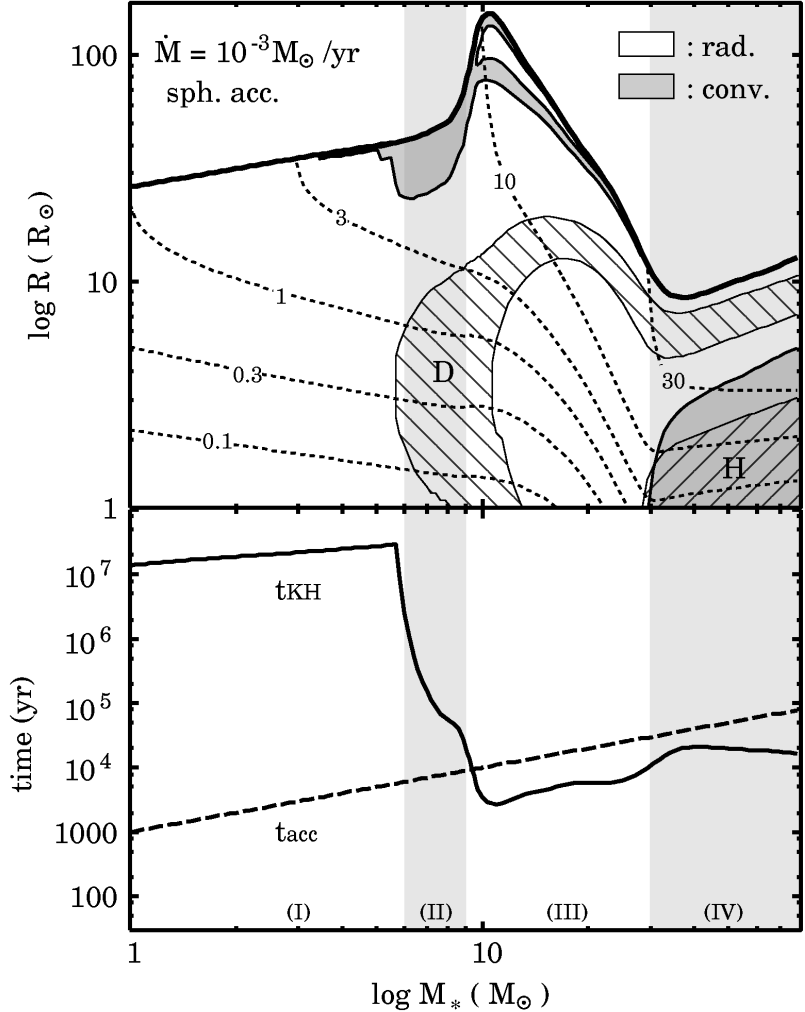}
\caption{Evolution of a protostar via spherical accretion 
at a rate $\dot{M}_\ast = 10^{-3}~M_{\odot}~{\rm yr}^{-1}$ 
(case MD3-S, taken from Paper I).
{\it Upper panel}: Evolution of the interior structure 
of the protostar. The gray-shaded areas denote convective layers.
The hatched areas indicate layers with active nuclear burning,
where the energy production rate exceeds 10\% of the steady rate
$0.1 L_{\rm D,st}/M_{\ast}$ for deuterium burning 
and $0.1 L_{\ast}/M_{\ast}$ for hydrogen burning. 
The thin dotted curves show the locations of mass coordinates
$M = 0.1$, 0.3, 1, 3, 10, and 30~$M_{\odot}$.
{\it Lower panel}: Evolution of the accretion timescale 
$t_{\rm acc}$ (dashed line), and Kelvin-Helmholz timescale 
$t_{\rm KH}$ (solid line).
In each panel the shaded background denotes the four evolutionary
phases; (I) adiabatic accretion, (II) swelling, (III) Kelvin-Helmholz
contraction, and (IV) main sequence accretion phases.}
\label{fig:str_log_1em3b1_sh}
\end{figure*}

Here, we briefly
review the evolution with spherically symmetric accretion studied in Paper I.
The upper panel of Figure \ref{fig:str_log_1em3b1_sh} shows the
evolution of the stellar radius and interior structure
at the accretion rate 
$\dot{M}_\ast = 10^{-3}~M_{\odot}~{\rm yr}^{-1}$ (case MD3-S).
We define the following four evolutionary phases from the
behavior of the stellar radius:
(I) gradual expansion ($M_\ast \lesssim 6~M_\odot$),
(II) swelling ($6~M_\odot \lesssim M_\ast \lesssim 10~M_\odot$),
(III) contraction ($10~M_\odot \lesssim M_\ast \lesssim 30~M_\odot$),
(IV) gradual expansion ($M_\ast \gtrsim 30~M_\odot$).
A key quantity to understand the variety of the evolution
is the ratio between the accretion timescale, 
\begin{equation}
t_{\rm acc} \equiv \frac{M_{\ast}}{\dot{M}_\ast} ,
\end{equation}
and the Kelvin-Helmholtz (KH) timescale,
\begin{equation}
\label{eq:tkh}
t_{\rm KH} \equiv \frac{G M_{\ast}^2}{R_{\ast} L_{\ast}} .
\end{equation}
The former is the evolutionary timescale, and the latter is
the timescale over which a star loses energy by radiation.
The lower panel of Figure \ref{fig:str_log_1em3b1_sh} displays
the evolution of these timescales.
We see that $t_{\rm KH}$ significantly decreases
during phase (II). As a result, the balance of these timescales changes
from $t_{\rm acc} \ll t_{\rm KH}$ in phase (I) to 
$t_{\rm acc} > t_{\rm KH}$ in phase (III) and (IV).
The decrease of $t_{\rm KH}$ is caused by a rapid increase of
the stellar luminosity $L_\ast$, which in turn is a result of the
decrease in opacity. In most of the stellar interior the opacity's
dependence on density and temperature is approximated by 
Kramers' law $\kappa \propto \rho T^{-3.5}$.
Thus, as the stellar mass increases, the temperature in the
interior rises and the opacity decreases.

\begin{figure*}
\plotone{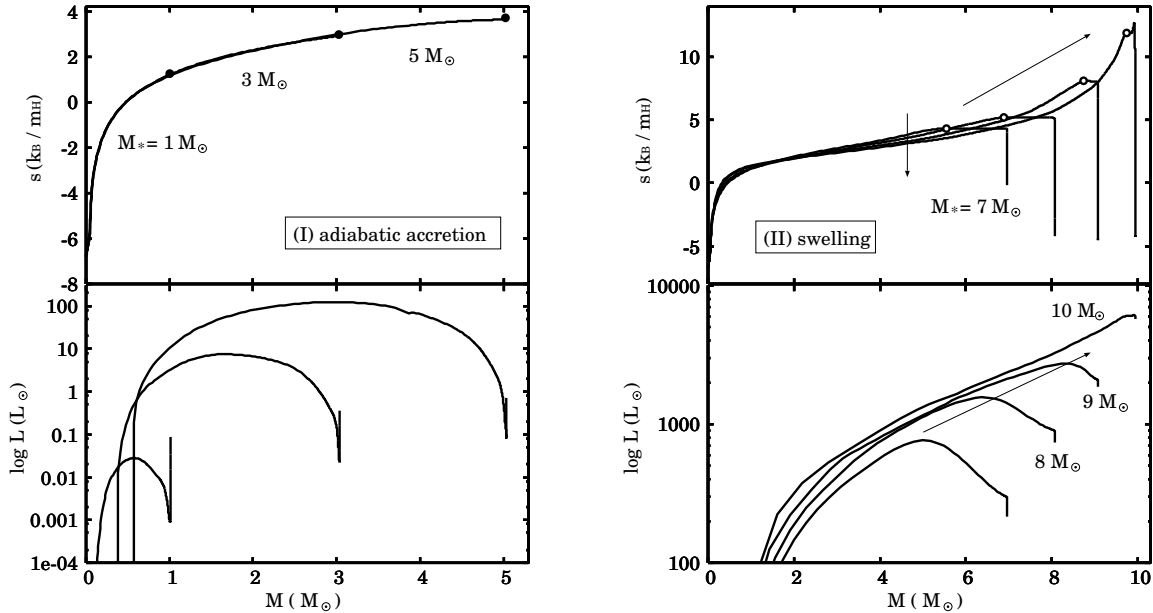}
\caption{ Radial profiles of the specific entropy 
and luminosity assuming spherical accretion at a rate 
$\dot{M}_\ast = 10^{-3}~M_{\odot}~{\rm yr}^{-1}$ 
(case MD3-S, taken from Paper I).
Each profile is shown as a function of the mass coordinate $M$.
The two panels display representative profiles
during the first two evolutionary
phases: (I) adiabatic accretion (left), and (II) swelling phase (right).
The filled circles on the entropy profiles in the left panel indicate
the post-shock values. The open circles in the right panel show
the values for the bottom edges of the convective layers.}
\label{fig:slpf_1em3_sh}
\end{figure*}

Figure \ref{fig:slpf_1em3_sh} shows snapshots of radial 
profiles of entropy and luminosity in the early phases
(I) and (II). 
At the assumed high accretion rate the entropy is initially efficiently 
brought into the stellar interior with minimal radiative loss
because of the high opacity during phase (I).
The instantaneous entropy profile simply traces 
the post-shock values from earlier times (adiabatic accretion). 
The average entropy within the star increases with accreted mass,
which according to equation (\ref{eq:r_entropy}) means that the
stellar radius should also grow.  This is indeed the case in 
phases (I) and (II).
As the average density decreases
and the temperature increases, the opacity decreases; 
radiative heat transport becomes more efficient with 
increasing stellar mass.

For $M_\ast \gtrsim 6~M_\odot$ (phase II) radiative
heat transport is efficient enough to modify the entropy distribution 
within the star.  The deep interior (where $\partial L/ \partial M > 0$) 
loses entropy, whereas the outer surface regions 
(where $\partial L/ \partial M < 0$) gain a significant 
amount of entropy.  This gain in entropy results in
the ``bloating up'' of the star up to $\gtrsim100~R_\odot$ at maximum. 
Figure \ref{fig:slpf_1em3_sh} shows that the boundary between
the heat-losing interior and heat-gaining outer layer, namely
where $\partial L/\partial M = 0$,
moves toward the stellar surface with increasing stellar mass.
Stahler, Palla \& Salpeter (1986) called this characteristic 
behavior of luminosity profiles a ``luminosity wave''.
The stellar surface luminosity $L_\ast$ rapidly increases 
when the luminosity wave approaches the stellar surface. 
After the luminosity wave passes through the surface, 
the star can efficiently lose energy by radiation. 
The star contracts to maintain 
virial equilibrium (Kelvin-Helmholz or KH contraction; phase III). 
The interior temperature rises during the contraction.
Active hydrogen burning begins when the central temperature 
exceeds $10^7$~K. After that, the stellar radius increases following the 
mass-radius relation of main sequence stars (phase IV).

\subsection{Case with Cold Disk Accretion}
\label{ssec:d_acc}

\subsubsection{Fiducial Model}
\label{sssec:fid}

\begin{figure*}
\epsscale{0.7}
\plotone{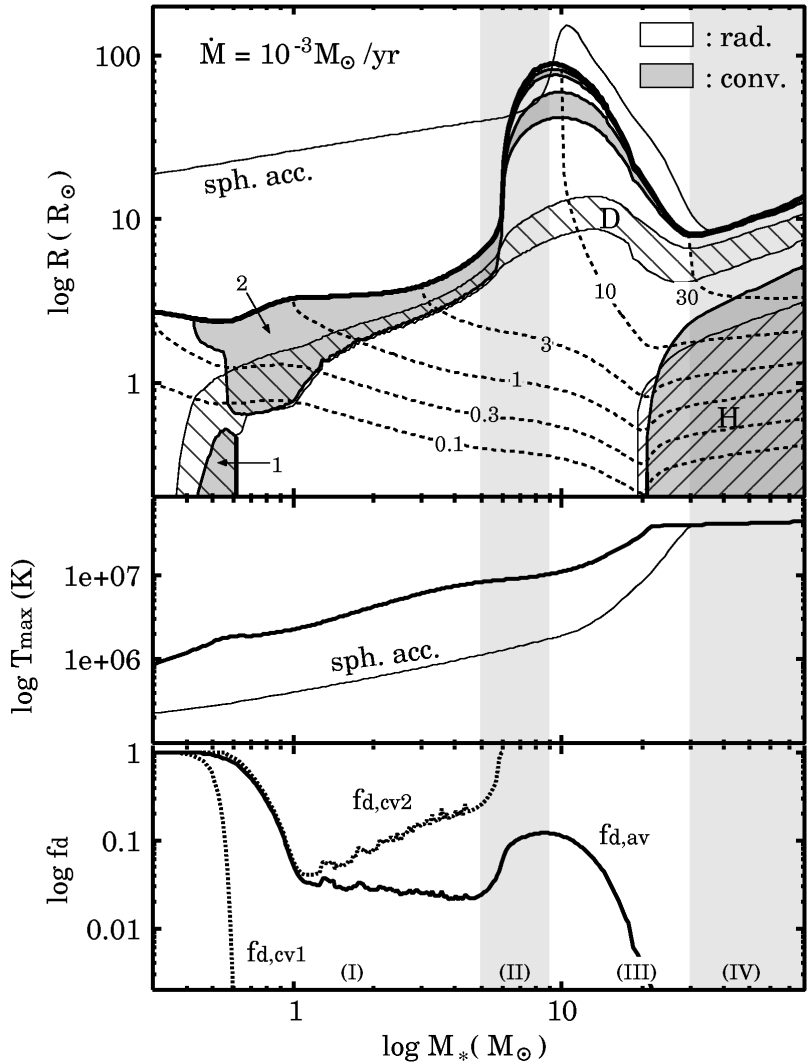}
\caption{Evolution of a protostar via disk accretion
at the rate $\dot{M}_\ast = 10^{-3}~M_{\odot}~{\rm yr}^{-1}$ 
(case MD3-D).
{\it Top panel} : Evolution of the interior structure
of a protostar. Illustration of the figure is the same as in
the upper panel of Fig. \ref{fig:str_log_1em3b1_sh}.
The thin solid line represents the evolution of the radius
via spherical accretion at the same rate 
(case MD3-S, taken from Paper I).
{\it Middle panel} :
Evolution of the maximum temperature within 
the star $T_{\rm max}$. Values for spherical accretion
at the same rate are plotted with a thin solid line.
In this figure the shaded background shows the four evolutionary
phases; (I) convection, (II) swelling, (III) Kelvin-Helmholz
contraction, and (IV) main sequence accretion phases.
{\it Bottom panel} :
Evolution of the deuterium concentrations of convective
layers $f_{\rm d, cv1/2}$ (dotted lines), and 
mass-averaged deuterium concentration,
$f_{\rm d, av}$ (solid line). }
\label{fig:struct_log_1em3b1}
\end{figure*}

We now consider stellar evolution with cold disk accretion 
at the same rate $\dot{M}_\ast = 10^{-3}~M_{\odot}~{\rm yr}^{-1}$ (case MD3-D).
First, we discuss the fiducial case with the $0.1~M_\odot$
initial model adopting $\beta = 1$ in equation (\ref{eq:s0}).
The top panel of Figure \ref{fig:struct_log_1em3b1} shows
the evolution of the stellar radius and interior structure.
We see that the evolution differs from the spherical 
accretion case for $M_\ast \lesssim 10~M_\odot$, as 
previously demonstrated by YB08
(also see Appendix \ref{ap:cp_yb} for comparison to YB08).
For $M_\ast \lesssim 5~M_\odot$ the stellar radius is much smaller 
than for spherical accretion. An outer convective zone appears 
in this phase. The protostar then abruptly inflates during the period
$5~M_\odot \lesssim M_\ast \lesssim 9~M_\odot$.
The maximum radius is $\simeq 90~R_\odot$ for $M_\ast \simeq 10~M_\odot$,
which is comparable to the results of the spherical accretion case. 
The stellar radius decreases for $M_\ast \gtrsim 10~M_\odot$, and then
finally follows the mass-radius relationship for main sequence stars when
$M_\ast \gtrsim 30~M_\odot$. The evolution for $M_\ast \gtrsim 10~M_\odot$
is quite similar to what we discussed above for spherical accretion. 

\begin{figure*}
\epsscale{0.7}
\plotone{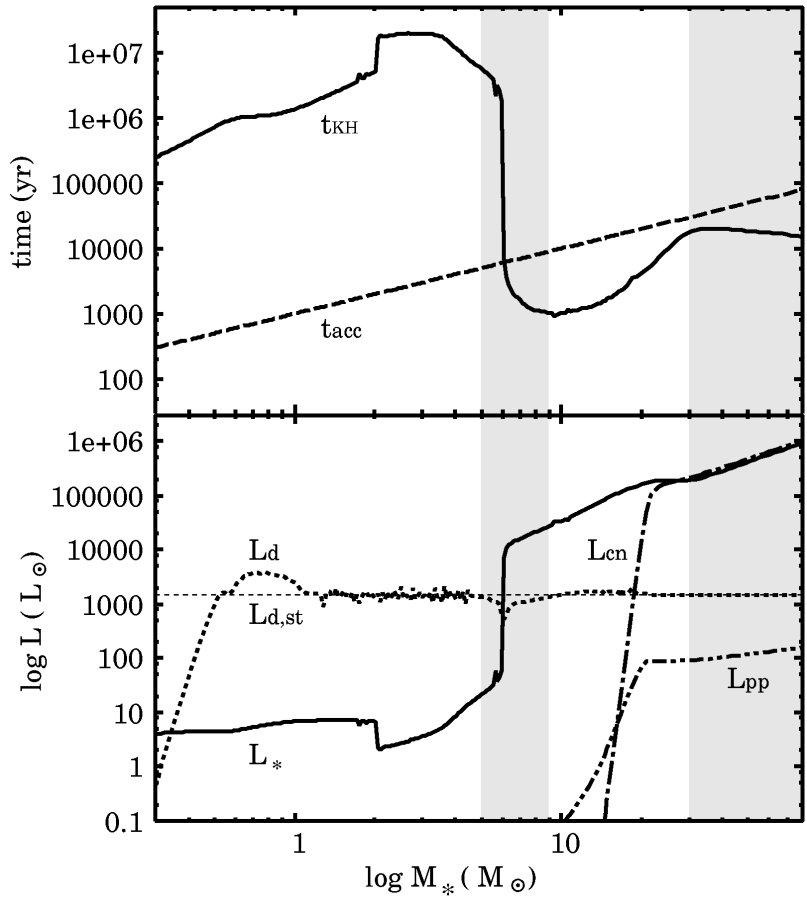}
\caption{
{\it Upper panel:} Evolution of the accretion timescale $t_{\rm acc}$
(dashed line) and Kelvin-Helmholz timescale $t_{\rm KH}$ (solid line).
{\it Lower panel:} 
Evolution of various contributions
to the luminosity of the protostar. The thick solid
line depicts the luminosity at the stellar surface $L_\ast$. 
The total energy production rate by each type of nuclear reaction 
is shown for 1) deuterium burning ($L_{\rm d}$, coarse dotted), 
2) pp-chain ($L_{\rm pp}$, dot-dot-dashed),
and 3) CNO-cycle ($L_{\rm CN}$, dot-dashed line). 
The thin dashed horizontal line indicates the steady 
deuterium burning rate $L_{\rm d,st}$.
In each panel the shaded background denotes the four evolutionary
phases as in Fig.\ref{fig:struct_log_1em3b1}.}
\label{fig:lev_tsc_1em3b1_ph}
\end{figure*}

We define the following four phases based on the 
evolutionary features: (I) convection ($M_\ast \lesssim 5~M_\odot$),
(II) swelling ($5~M_\odot \lesssim M_\ast \lesssim 9~M_\odot$),
(III) KH contraction ($9~M_\odot \lesssim M_\ast \lesssim 30~M_\odot$),
and (IV) main sequence accretion phase ($M_\ast \gtrsim 30~M_\odot$).
The top panel of Figure \ref{fig:lev_tsc_1em3b1_ph} shows 
the evolution of the accretion timescale and KH timescale.
We note that $t_{\rm KH}$ significantly decreases at 
$M_\ast \simeq 6~M_\odot$, when the protostar rapidly inflates.
The timescale balance sharply changes from 
$t_{\rm acc} \ll t_{\rm KH}$ to $t_{\rm acc} > t_{\rm KH}$ here.
As with spherical accretion, we attribute the sequence of the 
evolutionary phases from (I) to (III) to the inversion of the 
timescale balance. 
The lower panel of Figure \ref{fig:lev_tsc_1em3b1_ph} shows
that this change is due to the rapid increase of stellar 
luminosity $L_\ast$ caused by the decrease of opacity in the stellar
interior with increasing mass.
The detailed evolution in each phase is explained below.

\paragraph{Convection Phase}

The top panels of Figures \ref{fig:str_log_1em3b1_sh} and 
\ref{fig:struct_log_1em3b1} show that the  
evolution at $M_\ast \lesssim 5~M_\odot$ is quite different
between the spherical accretion case and disk accretion case.
In the disk accretion case the protostellar radius is about
one tenth of the value obtained for spherical accretion.
This is explained by equation (\ref{eq:r_entropy}) and the 
fact that the entropy content within the star is much lower
with disk accretion, a natural
consequence of the different accretion geometry.
With spherically symmetric accretion gas settles onto the star
with high entropy generated in the accretion shock front. 
For cold disk accretion, 
on the other hand, the entropy of the accreting gas
is reduced to the value in the stellar atmosphere 
by the gas' ability to radiate into free space.
Mass accretion hardly affects the average entropy 
in the stellar interior for the case of cold disk accretion.
Equation (\ref{eq:r_entropy}) shows that the stellar radius 
decreases according to $R_\ast \propto M_\ast^{-1/3}$ in this case.
We see this decrease for $M_\ast \lesssim 0.5~M_\odot$ 
in Figure \ref{fig:struct_log_1em3b1}.

\begin{figure*}
\epsscale{0.7}
\plotone{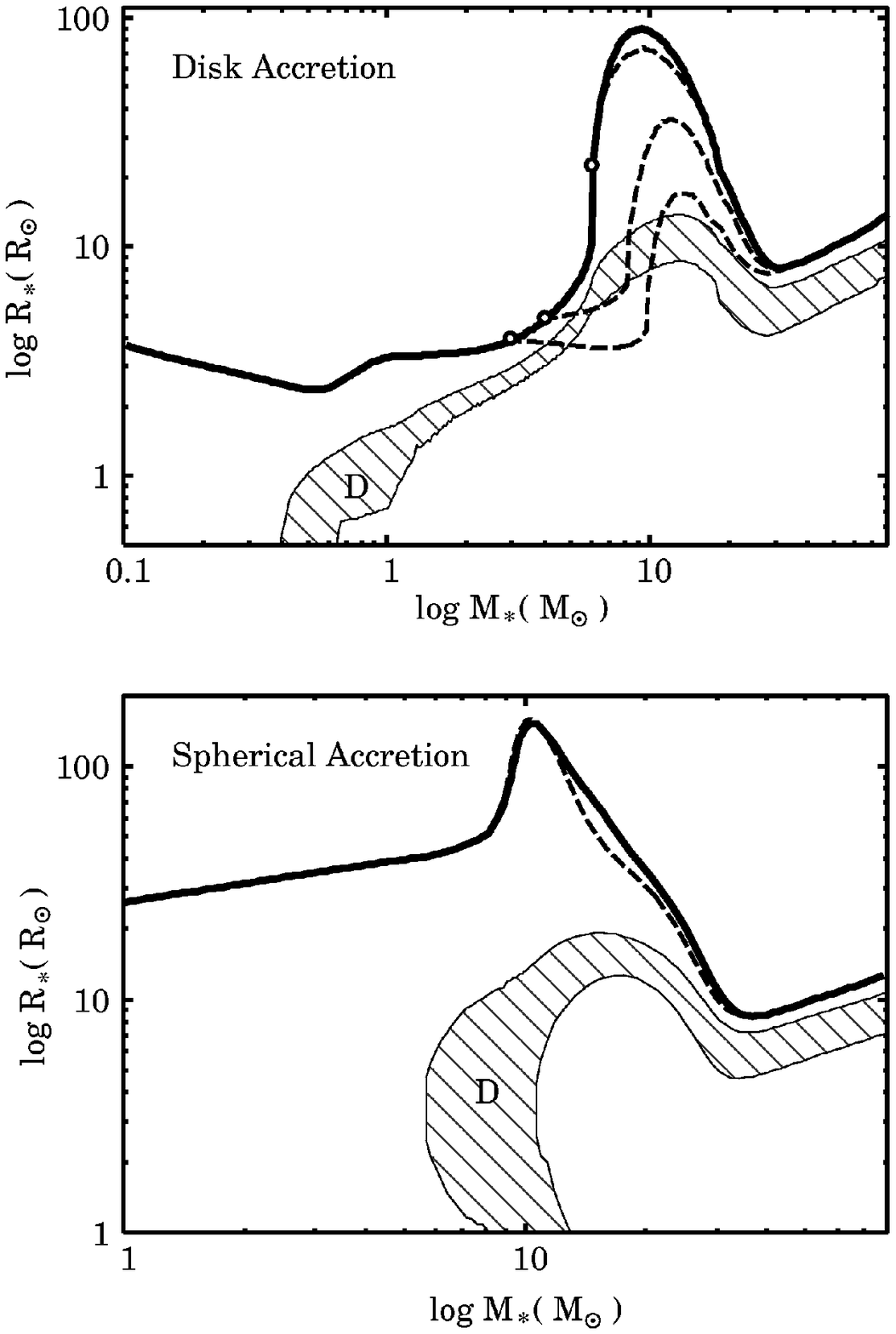}
\caption{Effects of deuterium burning on protostellar evolution.
The upper and lower panels show the evolution 
via disk and spherical accretion for 
$\dot{M}_\ast = 10^{-3}~M_{\odot}~{\rm yr}^{-1}$.
In both panels the thick solid line represents the stellar radius.
The hatched area denotes a layer with active deuterium
burning, where the energy production rate is larger than 
$0.1 L_{\rm D,st}/M_\ast$. 
In the upper panel the dashed lines show the
evolution of the stellar radius after turning off deuterium 
burning at 3, 4, and $6~M_\odot$, respectively, 
which are indicated by open circles.
In the lower panel the dashed line shows the evolution 
with no deuterium burning from the beginning.}
\label{fig:mr_doff}
\end{figure*}

Figure \ref{fig:struct_log_1em3b1} shows that deuterium burning 
(D-burning) begins when $M_\ast \simeq 0.4~M_\odot$.
This is in stark contrast to the spherical accretion case, where
it begins much later ($M_\ast \simeq 6~M_\odot$), as shown in
Figure \ref{fig:mr_doff}. 
This difference reflects the different evolution of the 
maximum temperature within the protostar $T_{\rm max}$
(middle panel of Fig. \ref{fig:struct_log_1em3b1}).
The early rise of $T_{\rm max}$ with disk accretion is
explained by the relation, 
\begin{equation}
T = \frac{\mu}{\cal{R}} \frac{P}{\rho} 
  \propto \frac{M_{\ast}}{R_{\ast}},
\label{eq:t_typ}
\end{equation}
where $T$ is the typical temperature within the star
${\cal{R}}$ is the gas constant, and $\mu$ is the mean molecular weight.
Equations (\ref{eq:typ_rhop}) were used again to derive 
the dependence on $M_\ast / R_\ast$.
Equation (\ref{eq:t_typ}) means that, at the same stellar mass, 
the interior temperature is higher for the smaller stellar radius.
Since the stellar radius $R_\ast$ is smaller for the disk accretion,
the central temperature is higher for the same stellar mass. 
Thus, D-burning begins much earlier in the disk accretion case. 

Figure \ref{fig:struct_log_1em3b1} shows that convective zones 
emerge soon after deuterium burning begins. First, a convective 
zone appears near the center for $M_\ast \simeq 0.5~M_\odot$.
However, this zone disappears soon, when available deuterium
is exhausted.
The D-burning layer moves to the outer part of the star, where
fresh deuterium still remains. 
Another convective zone forms there, connecting the D-burning 
layer to the stellar surface, where newly accreted deuterium can
be mixed down to the D-burning layer. 
When $M_\ast \gtrsim 1~M_\odot$ the total energy production rate 
by D-burning becomes nearly equal to the steady-state rate
(bottom panel of Fig. \ref{fig:lev_tsc_1em3b1_ph}), 
\begin{equation}
L_{\rm D, st} \equiv \dot{M}_\ast \delta_{\rm D} = 
                1500~L_{\odot} \left( 
                               \frac{\dot{M}_\ast}{10^{-3}~M_{\odot}/{\rm yr}} 
                             \right)
                             \left(
                               \frac{[ {\rm D/H} ]}{2.5 \times 10^{-5}} 
                             \right), 
\label{eq:l_dst}
\end{equation}
where $\delta_{\rm D}$ is the energy available from
the deuterium burning per unit gas mass.

\begin{figure*}
\plotone{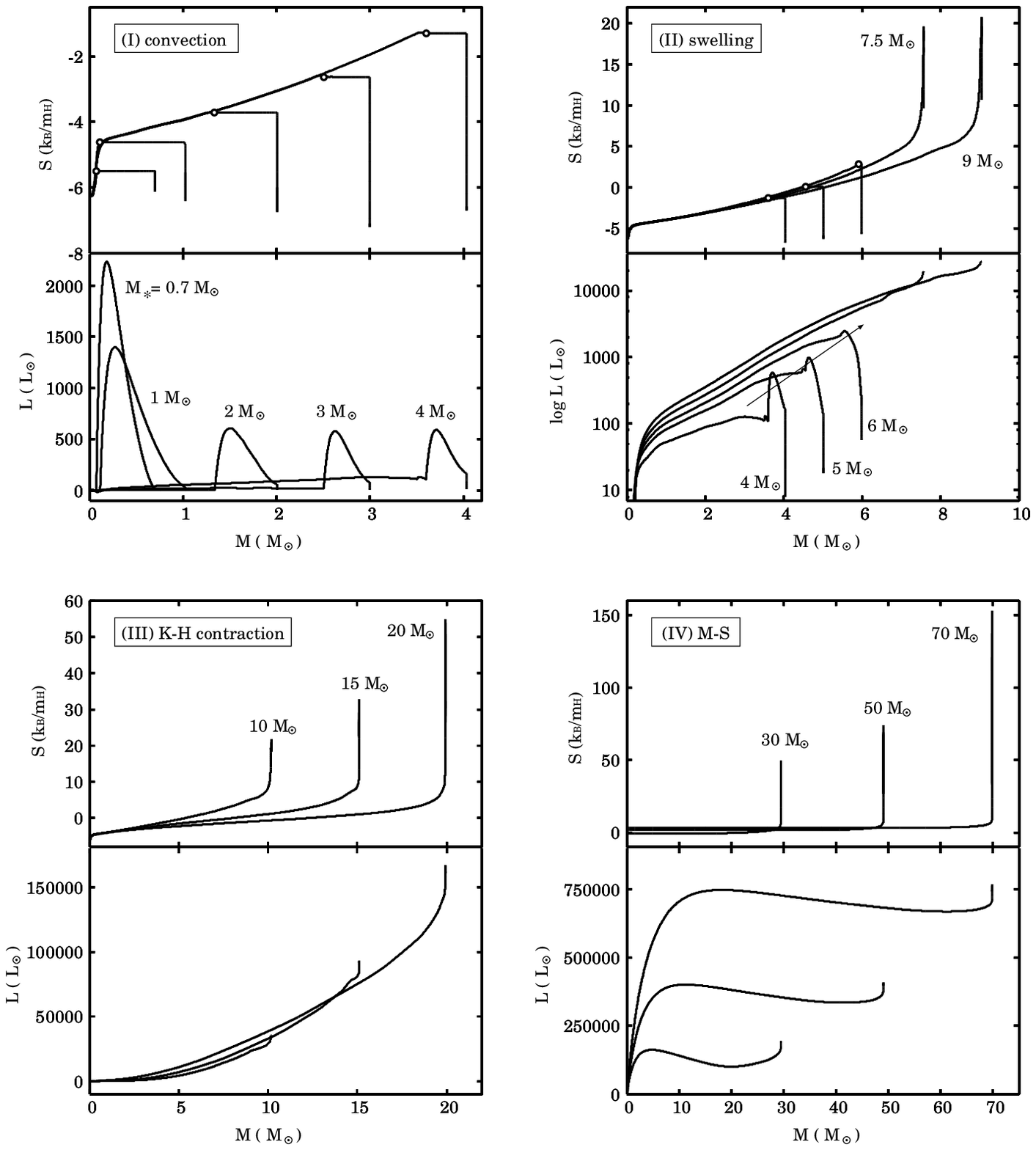}
\caption{Radial profiles of the specific entropy 
and luminosity for disk accretion at the rate 
$\dot{M}_\ast = 10^{-3}~M_{\odot}~{\rm yr}^{-1}$ (case MD3-D).
Illustration of the figure is the same as in Fig. \ref{fig:slpf_1em3_sh}.
The four panels correspond to the four evolutionary phases, 
(I) convection (upper left), (II) swelling 
(upper right), (III) Kelvin-Helmholtz contraction (lower left), 
and (IV) main sequence accretion (lower right) phases.
In the upper panels the open circles indicate the bottom edges of 
convective layers.}
\label{fig:sl_1em3_beta1_ph}
\end{figure*}

For $M_\ast \gtrsim 1~M_\odot$, the radiative core extends outward
and the outer convective layer with its D-burning shell becomes 
geometrically thinner. 
This evolution is a result of the decrease of opacity in 
the stellar interior; 
radiative heat transport becomes more efficient with 
the lower opacity. At the end of phase (I)
convection ceases where the radiative heat transport
is efficient enough to carry away the heat generated by
deuterium burning. 
Fresh deuterium is no longer supplied to the D-burning layer 
which has now become radiative.

Figure \ref{fig:struct_log_1em3b1} shows that the stellar radius
gradually increases after the ignition of deuterium burning.
This is because deuterium burning enhances the average entropy
in the stellar interior.
The surface convective layer absorbs the entropy generated 
by deuterium burning.
Figure \ref{fig:sl_1em3_beta1_ph}-(I) clearly shows that the entropy 
in the surface convective layer rises with increasing stellar mass.
The effect of deuterium burning is also seen in
Figure \ref{fig:mr_doff}, which depicts the evolution of the
stellar radius when deuterium burning is artificially turned
off at some point (dashed lines). 
We see that, if deuterium burning is turned off at
$M_\ast = 3~M_\odot$, the stellar radius ceases to increase.

Figure \ref{fig:mr_doff} also presents the evolution 
via spherical accretion without deuterium burning (dashed line).
We see that, unlike the case of disk accretion explained above, 
deuterium burning hardly affects the evolution. 
A key quantity to interpret this difference is the opacity
when deuterium burning begins. 
With spherical accretion, deuterium burning begins late, when
$M_\ast \simeq 6~M_\odot$ and the opacity within the star
has already decreased sufficiently.
The radiative heat transport is so efficient that the heat 
generated by deuterium burning is carried away before the
convection arises.
With disk accretion, on the other hand, the opacity is still
high at the ignition of deuterium burning when $M_\ast \simeq 0.3~M_\odot$.
Deuterium burning initiates the convection due to the inefficient radiative
heat transport. In this case deuterium burning affects the stellar evolution
until opacity becomes low enough owing to increasing the stellar mass.

\paragraph{Swelling Phase}

The stellar radius rapidly increases in this phase.
However, this is not a homologous expansion of the star.
Only a small percent of the mass near the surface significantly bloats up. 
For $M_\ast \simeq 10~M_\odot$, for example, 10\% of the total stellar mass
fills 90\% of the radius and 99.9\% of the volume.  
Figure \ref{fig:struct_log_1em3b1} also shows that the stellar 
interior structure changes during the swelling.
The surface convective zone leaves the D-burning layer
and moves outward for $M_\ast \simeq 6~M_\odot$. 
The radiative core covers most of the
stellar interior.

What is the mechanism of the significant swelling?
We contend that deuterium burning cannot be the primary cause, 
as demonstrated in the top panel of Figure \ref{fig:mr_doff}. 
We see that, even if we completely (artificially) 
turn off deuterium burning, the 
protostar subsequently inflates in all examined cases.  
Deuterium burning enhances the swelling, however;
the earlier deuterium burning is turned off, the later
the swelling occurs with a smaller maximum radius. 

\begin{figure*}
\epsscale{0.7}
\plotone{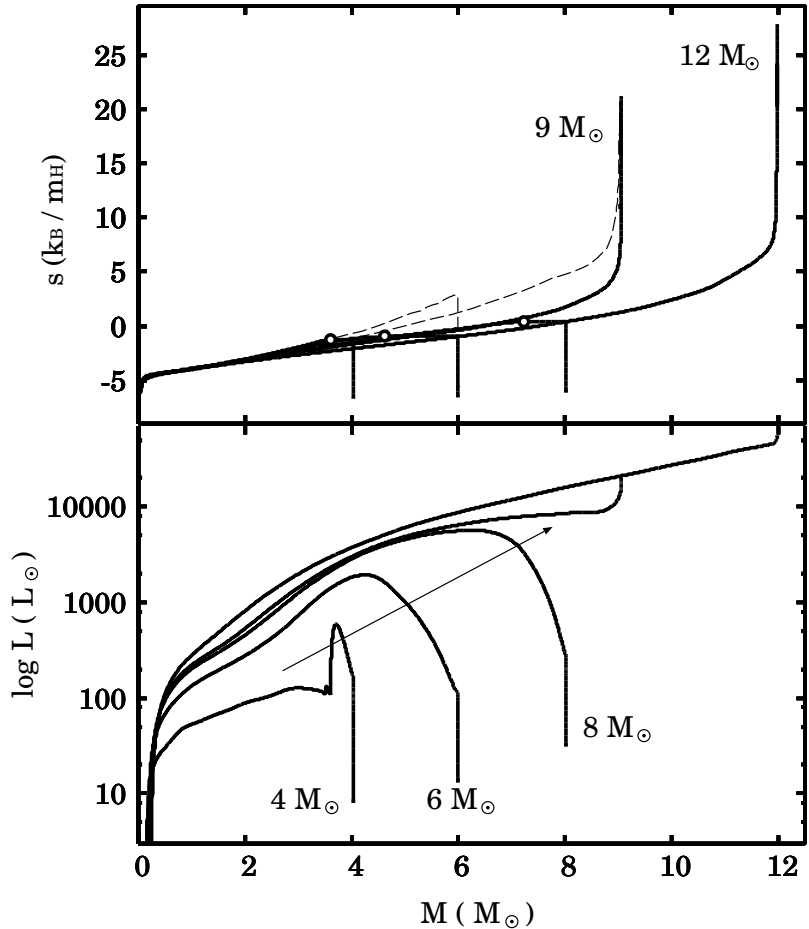}
\caption{Evolution of radial profiles of specific entropy
and luminosity after turning off deuterium burning at the point when
$M_\ast = 4~M_\odot$. 
The lines labeled $4~M_\odot$ show the profiles just before
turning off deuterium burning.
The thin dashed lines in the upper panel show the entropy
profiles for $M_\ast = 6~M_\odot$ and $9~M_\odot$ in the fiducial
case with deuterium burning.}
\label{fig:slpf_1em3_ph_doff4}
\end{figure*}

We examine the mechanism of the swelling by considering the evolution
after turning off deuterium burning. 
Figure \ref{fig:slpf_1em3_ph_doff4} shows snapshots of entropy 
and luminosity profiles in each case. 
For $M_\ast \lesssim 8~M_\odot$ we find the characteristic feature
of the ``luminosity wave'' in the luminosity profiles as in the
spherical accretion case (see Section~\ref{ssec:sp_acc},  
and Fig. \ref{fig:slpf_1em3_sh}). 
The heat is transported from the interior where 
$\partial L / \partial M > 0$ to the surface layer where 
$\partial L / \partial M < 0$.
The maximum luminosity increases with increasing stellar mass. 
At the same time the heat-losing interior extends outward toward 
the stellar surface. 
This characteristic evolution is caused by the decrease of opacity
within the star. More and more entropy embedded in the deep 
stellar interior is transferred outward by radiation 
as the opacity decreases. 
The surface layer bloats up, because it has a high entropy through
absorbing part of the transported entropy.
Figure \ref{fig:slpf_1em3_ph_doff4} shows that the luminosity wave
reaches the stellar surface for $M_\ast \simeq 9~M_\odot$. 
After that, $\partial L / \partial M > 0$ throughout the
star and there is no entropy-receiving layer. 
However, the swelling still continues with disk accretion.  
This is because the newly accreted gas 
settles onto the star with the same high entropy 
of the stellar atmosphere.
As a result, the surface layer always has high entropy.
On the other hand, the stellar interior continues to lose 
entropy with decreasing opacity. The significant entropy
loss in the stellar interior finally overcomes the swelling,
and the star begins a phase of KH contraction.

Evolution of luminosity and entropy profiles with deuterium burning
is shown in Figure \ref{fig:slpf_1em3_sh}.
The evolution is qualitatively similar to that without deuterium
burning presented in Figure \ref{fig:slpf_1em3_ph_doff4}.
Deuterium burning increases the entropy within the star, which makes 
the heat transport after the opacity decreases more significant. 
Therefore, deuterium burning enhances the swelling.
Figure \ref{fig:slpf_1em3_sh} shows that with deuterium burning
the luminosity wave reaches 
the surface when $M_\ast \simeq 6~M_\odot$. 
The stellar surface luminosity $L_\ast$ significantly increases 
at this moment, which causes the inversion of the timescale balance 
(Fig. \ref{fig:lev_tsc_1em3b1_ph}).

\paragraph{Later Phases}

For $M_\ast \gtrsim 9~M_\odot$ the protostar begins to contract
($t_{\rm KH} < t_{\rm acc}$). 
The protostar efficiently loses its energy by radiation. 
Figure \ref{fig:sl_1em3_beta1_ph} shows that the entropy in the 
stellar interior decreases during this phase. 
The maximum temperature within the star soon exceeds $10^7$~K.
Hydrogen burning begins first via the pp-chain followed by 
CNO-cycle reactions, which lead to a 
convective core emerges when $M_\ast \simeq 20~M_\odot$. 
The stellar surface luminosity becomes nearly equal to the total
energy production rate by CNO-cycle hydrogen burning for
$M_\ast \simeq 30~M_\odot$ (Fig. \ref{fig:lev_tsc_1em3b1_ph}). 
The stellar mass when the protostar reaches the ZAMS 
$M_{\ast, {\rm ZAMS}}$ is analytically estimated by equating
the accretion timescale to the KH timescale of ZAMS stars:
\begin{equation}
t_{\rm acc} \simeq t_{\rm KH,ZAMS} .
\label{eq:t_mzams}
\end{equation}
For $10~M_\odot \lesssim M_\ast \lesssim 100~M_\odot$ 
the mass-radius and mass-luminosity relationships of ZAMS stars 
(e.g. Schaller et al. 1992) are well fitted by:
\begin{equation}
L_{\ast, {\rm ZAMS}} = 1.4 \times 10^4~L_\odot 
\left( \frac{M_\ast}{10~M_\odot} \right)^2 ,
\label{eq:l_zams}
\end{equation}
\begin{equation}
R_{\ast, {\rm ZAMS}} = 3.9~R_\odot 
\left( \frac{M_\ast}{10~M_\odot} \right)^{0.55} .
\label{eq:r_zams}
\end{equation}
Substituting these relations into equation (\ref{eq:t_mzams}), we get
\begin{equation}
M_{\ast, {\rm ZAMS}} \simeq 30.5~M_\odot
\left( \frac{\dot{M}_\ast}{10^{-3}~M_{\odot}~{\rm yr}^{-1}}
\right)^{0.645} ,
\label{eq:mzams}
\end{equation}
which agrees well with our calculations.
Since the above argument is independent of different accretion geometries,
so is $M_{\ast, {\rm ZAMS}}$.

\subsubsection{Dependence on Initial Models}
\label{sssec:imod}

\begin{figure*}
\epsscale{0.7}
\plotone{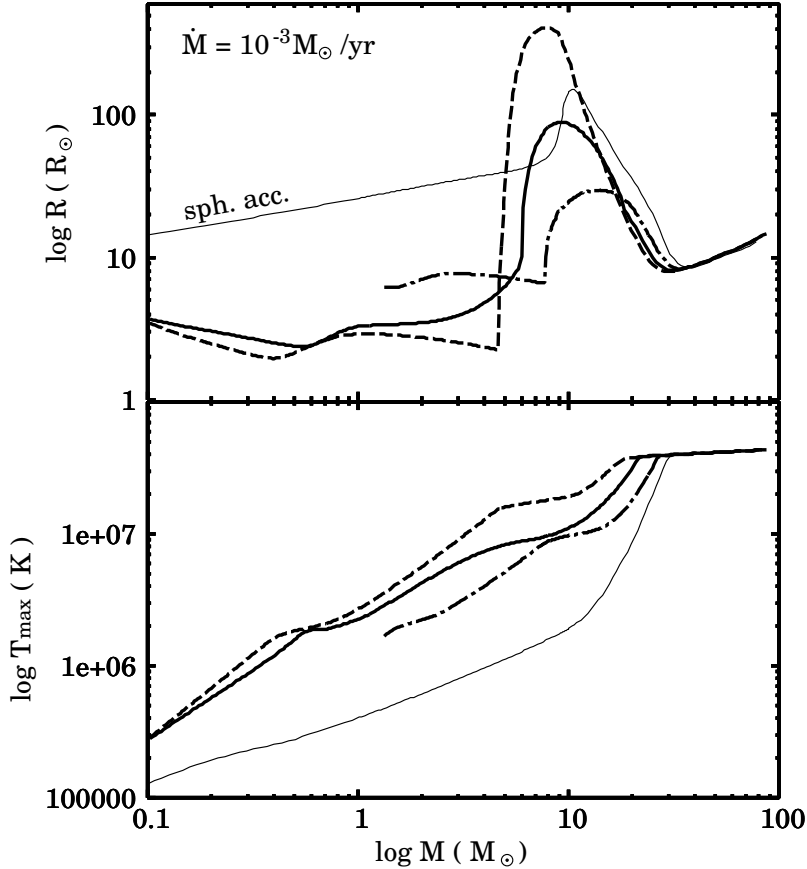}
\caption{Variation of evolution via disk accretion with 
different initial models with an adopted accretion rate
$\dot{M}_\ast = 10^{-3}~M_{\odot}~{\rm yr}^{-1}$.
The upper and lower panels display the evolution of the
protostellar radius and maximum temperature within the star. 
In both panels the solid line shows the evolution in the fiducial case
explained in Section \ref{sssec:fid} (case MD3-D), the dashed 
line represents the evolution assuming a shallower initial entropy 
distribution (case MD3-D-b0.1), and the
dot-dashed line shows the evolution with a fully
convective initial model (case MD3-cv).
For comparison we also display the evolution for spherical accretion 
at the same accretion rate (case MD3-S) with a thin solid line.   }
\label{fig:1em3_ph_ini}
\end{figure*}

Next, we examine how the protostellar evolution changes
with different initial models. 
To this end we consider two initial models that differ from the
the fiducial one.
One is a $0.1~M_\odot$ star with a shallower initial entropy 
distribution ($\beta = 0.1$ in equation (\ref{eq:s0}); 
case MD3-D-b0.1). 
The other is a fully convective $1.3~M_\odot$ star 
(case MD3-D-cv), which was used by Palla \& Stahler (1992) to calculate
the evolution at $\dot{M}_\ast = 10^{-4}~M_{\odot}~{\rm yr}^{-1}$ 
via cold disk accretion.
Figure \ref{fig:1em3_ph_ini} shows the evolution of stellar radius
and maximum temperature in these cases. 
The basic evolution is similar.
A protostar undergoes swelling and contraction, finally 
reaching the main sequence accretion phase when $M_\ast \simeq 30~M_\odot$. 
However, the evolution differs quantitatively for the different initial models.
For example the maximum stellar radius is $\simeq 400~R_\odot$ 
for case MD3-D-b0.1, but $\simeq 30_\odot$ for case MD3-D-cv 
(the evolution of the interior structure in these cases
also differs slightly from that for case MD3-D described 
in Section~\ref{ssec:d_acc}; see Appendix~\ref{ap:b0.1} for details).

This is in stark contrast to the spherical accretion case, where
the evolution is almost independent of different initial models.
For spherically symmetric accretion the evolution of stellar radius in the 
adiabatic accretion phase is well approximated by the relation
(Stahler, Palla \& Salpeter 1986):
\begin{equation}
R_{\ast} \simeq  26~R_{\odot} \left( \frac{M_{\ast}}{M_{\odot}} \right)^{0.27} 
                 \left( \frac{\dot{M}_\ast}
                             {10^{-3}~M_{\odot}~{\rm yr}^{-1}} \right)^{0.41}.
\label{eq:r_sps86}
\end{equation}
Our calculated evolution with spherical accretion (case MD3-S) also
obeys this relation before the swelling occurs.
Stahler, Palla \& Salpeter (1986) showed that the evolution of the radius
converges to the above relation even beginning with different initial
models.  
With spherical accretion entropy generated at an accretion shock
front is taken into the stellar interior along with the accreted gas. 
If the initial radius is too large compared to equation 
(\ref{eq:r_sps86}), the accretion shock front is 
weak due to the shallow gravitational potential well. 
The entropy produced at the accretion shock front
is reduced, which decreases average entropy within the star.
As a result the protostar contracts. 
If the initial radius is too small, the opposite occurs, increasing
the radius. 
With disk accretion, however, this regulation process does not operate.
The entropy of the accreted material is set to the value in the
stellar atmosphere. Since the structure of the stellar atmosphere
differs with different initial models, the subsequent evolution also
changes (also see Hartmann et al. 1997 for low-mass protostars).

\begin{figure*}
\epsscale{0.7}
\plotone{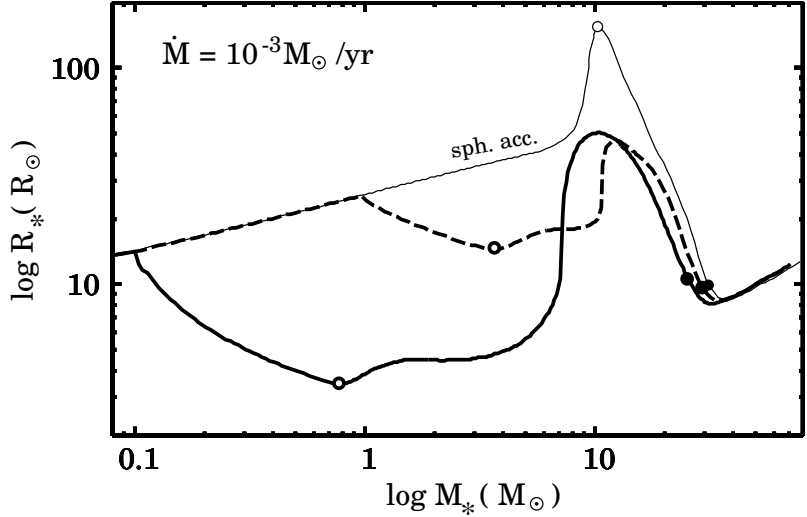}
\caption{Evolution of the stellar radius when the accretion geometry is
switched from spherical accretion to disk accretion at a constant rate
$\dot{M}_\ast = 10^{-3}~M_{\odot}~{\rm yr}^{-1}$.
The thick solid and dashed lines show the evolution when the
geometry is changed at $M_\ast = 0.1~M_\odot$ and $1~M_\odot$
(cases MD3-SDm0.1 and MD3-SDm1), respectively.
The open circles mark the epoch when the total energy production rate
by deuterium burning reaches 80\% of the steady 
burning rate $L_{\rm D,st}$.
The filled circles indicate the epoch when the total energy 
production rate by hydrogen burning exceeds 80\% of the luminosity
at the stellar surface $L_\ast$.
The thin line show the evolution assuming spherical accretion only 
(case MD3-S). }
\label{fig:m_rsh}
\end{figure*}

How should we choose the initial model? We infer this by considering 
the evolution of accretion flow toward the protostar.
Just after a protostar is born, the protostar gathers material 
from its immediate vicinity with low angular momentum. 
The geometry of the accretion flow is nearly spherical in this early phase.
Afterwards, gas originally far from the star falls toward the
star with increasingly higher angular momentum.
A circumstellar disk forms, and 
the protostar continues to grow via disk accretion.
The protostellar evolution in this scenario is calculated
by switching the accretion geometry from spherical to disk 
accretion at some point. 

Figure \ref{fig:m_rsh} shows some examples of protostellar 
evolution, whereby the accretion geometry is switched for $M_\ast = 0.1~M_\odot$ 
and $1~M_\odot$. The protostellar radius increases
in the early spherical accretion phase according to equation
(\ref{eq:r_sps86}).
The radius subsequently decreases after the geometry is switched to cold disk accretion,
because the accreting gas brings lower entropy into
the stellar interior.
The interior temperature increases as the protostar contracts.
Active deuterium burning begins when the maximum temperature exceeds
$\simeq 10^6$~K.
Figure \ref{fig:m_rsh} shows that, the earlier the geometry is switched, 
the earlier deuterium burning begins.
The subsequent evolution is similar to that in our fiducial case
MD3-D explained in Section \ref{ssec:d_acc}.
The maximum radius is $\simeq 50~R_\odot$ in both examined cases.
The protostar reaches the main sequence accretion phase at
$M_\ast \simeq 30~M_\odot$, which is independent of the history
of accretion geometry.

\subsection{Comparison with Primordial Protostars}
\label{ssec:metal}

\begin{figure*}
\epsscale{0.7}
\plotone{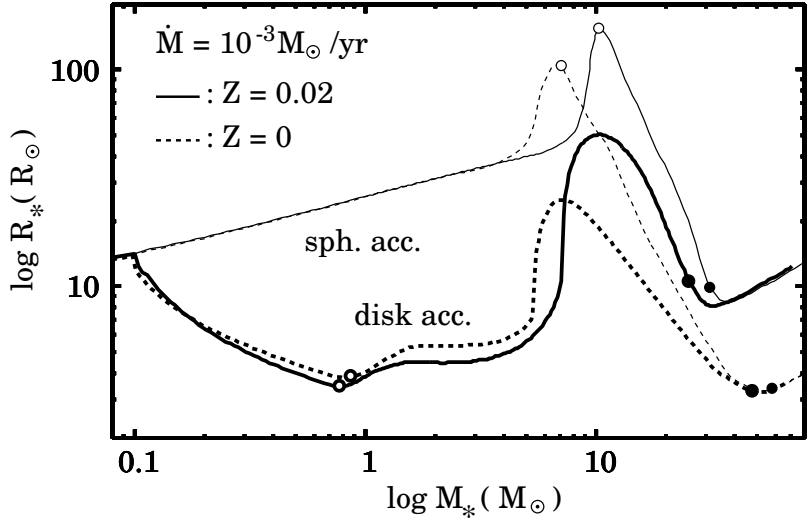}
\caption{Comparison with primordial cases:
Evolution of protostellar radius for an accretion rate 
$\dot{M} = 10^{-3}~M_{\odot}~{\rm yr}^{-1}$.
The thick solid and dotted lines show the evolution 
via disk accretion at $Z = 0.02$ (case MD3-SDm0.1) and 
$Z = 0$ (case MD3-SDm0.1-z0), respectively. 
The accretion geometry was switched from 
spherical accretion to disk accretion when $M_\ast = 0.1~M_\odot$.
The thin solid and dashed lines show the corresponding
evolution via spherical accretion 
for each metallicity (cases MD3-S and MD3-S-z0).
The meanings of the open and filled circles on the curves are
the same as in Figure \ref{fig:m_rsh}.}
\label{fig:m_rz}
\end{figure*}

Rapid mass accretion 
$\dot{M}_\ast \sim 10^{-3}~M_{\odot}~{\rm yr}^{-1}$ is also expected 
during the formation of primordial ($Z = 0$) stars. 
Here, we compare the evolution of primordial protostars to
that of present-day ($Z = Z_\odot$) massive protostars. 
Figure \ref{fig:m_rz} presents the evolution of the stellar radius 
with metallicities $Z = 0$ (case MD3-SDm0.1-z0) and
$Z = Z_\odot$ (case MD3-SDm0.1).
In these cases the accretion geometry is switched from spherical
accretion to disk accretion at $M_\ast = 0.1~M_\odot$. 
The basic evolution with $Z = 0$ is the same as in the 
present-day case. 
After deuterium burning begins for $M_\ast \simeq 0.8~M_\odot$, 
the protostar undergoes the four evolutionary phases of 
convection, swelling, KH contraction, and main sequence accretion. 

However, Figure \ref{fig:m_rz} also shows some quantitative differences
between the primordial and present-day cases.
At $Z = 0$, for example, the swelling occurs for $M_\ast \simeq 6~M_\odot$,
which is slightly earlier than for solar metalicities.
A similar difference is also found with spherical accretion, 
as shown in Figure \ref{fig:m_rz}. 
These differences are attributed to the fact that the opacity is 
lower at lower metallicity in the same thermal state. 
As explained in Section \ref{ssec:sp_acc} and \ref{ssec:d_acc}, 
the swelling occurs as a result of the decrease of opacity 
in the stellar interior with increasing the stellar mass. 
The swelling occurs earlier at $Z=0$, because the opacity
decreases sufficiently at an earlier time.
 
Another difference between the primordial and present-day case 
is the stellar radius in the main sequence accretion phase;
the ZAMS radius for $Z=0$ is much smaller than that for $Z = Z_\odot$,
independent of the assumed geometry.
In the primordial case the initial abundance of C, N, and O nuclei,
necessary catalysts for the CNO-cycle hydrogen burning,
is zero. CNO-cycle reactions do not begin
until these atoms are provided by helium triple-$\alpha$ burning,
which occurs at $T \simeq 10^8$~K.  Because
pp-chain hydrogen burning does not supply sufficient energy to 
halt the star's KH contraction, the pre-main contraction phase 
continues up to the higher central 
densities and temperatures that allow triple-$\alpha$ reactions.
This ZAMS star thus has a much smaller radius than its solar 
metallicity counterpart.

\section{Dependence on Mass Accretion Rates}
\label{sec:mdot_dep}

\begin{figure*}
\epsscale{0.7}
\plotone{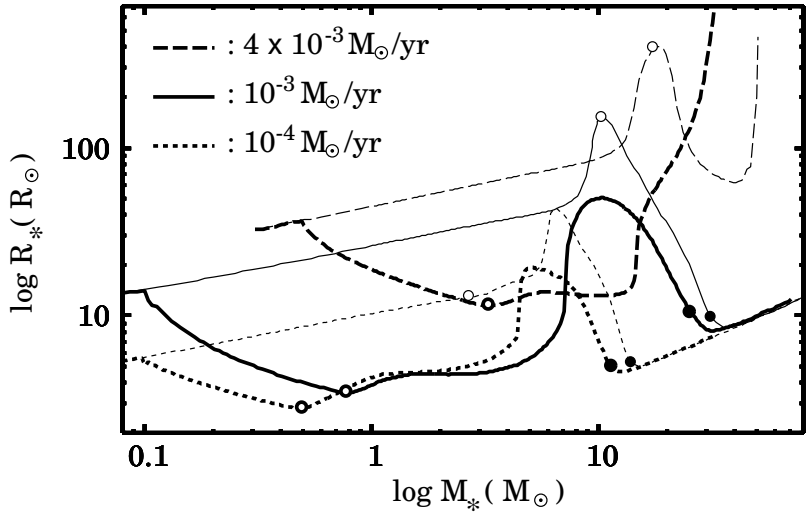}
\caption{Evolution of the protostellar radius with disk accretion 
at different accretion rates $\dot{M} = 10^{-4}~M_{\odot}~{\rm yr}^{-1}$ 
(case MD4-SDm0.1, dotted line), $10^{-3}~M_{\odot}~{\rm yr}^{-1}$ 
(case MD3-SDm0.1, solid line), and 
$4 \times 10^{-3}~M_{\odot}~{\rm yr}^{-1}$ (case MD4x3-SDm0.5, dashed line).
The evolution with spherical accretion at each rate is also plotted with
thin lines for comparison (cases MD4-S, MD3-S, and MD4x3-S).
The meanings of the open and filled circles on the curves are
the same as in Figure \ref{fig:m_rsh}.}
\label{fig:m_r_mdotv}
\end{figure*}

Finally, we examine how the protostellar evolution changes
with different accretion rates.  
As pointed out in Section \ref{sssec:imod}, the evolution 
quantitatively varies with different initial models even 
at the same accretion rate. 
In this section we focus on the cases where the accretion
geometry evolves from spherical accretion to disk accretion
in early phases. Figure \ref{fig:m_r_mdotv} displays the evolution of 
the stellar radius at the accretion rates of 
$\dot{M}_\ast = 10^{-4}~M_{\odot}~{\rm yr}^{-1}$ (case MD4-SDm0.1),
$10^{-3}~M_{\odot}~{\rm yr}^{-1}$ (case MD3-SDm0.1), and 
$4 \times 10^{-3}~M_{\odot}~{\rm yr}^{-1}$ (case MD4x3-SDm0.5).
With the rate $10^{-4}~M_{\odot}~{\rm yr}^{-1}$, the protostar
undergoes the same four evolutionary phases as with the rate
$10^{-3}~M_{\odot}~{\rm yr}^{-1}$. 
The maximum radius decreases with decreasing
the accretion rate. The same dependence is also seen with spherical
accretion. Note that, however, the maximum radius varies with 
different initial models under the assumption of a freely radiating 
photospheric boundary ("disk accretion")
as explained in Section \ref{sssec:imod}.
Figure \ref{fig:m_r_mdotv} also shows that, at lower accretion rates,
the protostar arrives to the main-sequence accretion phase at a lower
stellar mass. We argue that this dependence is robust.
The stellar mass at the arrival to the ZAMS is just 
determined by $t_{\rm acc} \simeq t_{\rm KH, ZAMS}$ as shown 
by equation (\ref{eq:mzams}), which is independent of different 
accretion geometries and different initial models.

\begin{figure*}
\epsscale{0.7}
\plotone{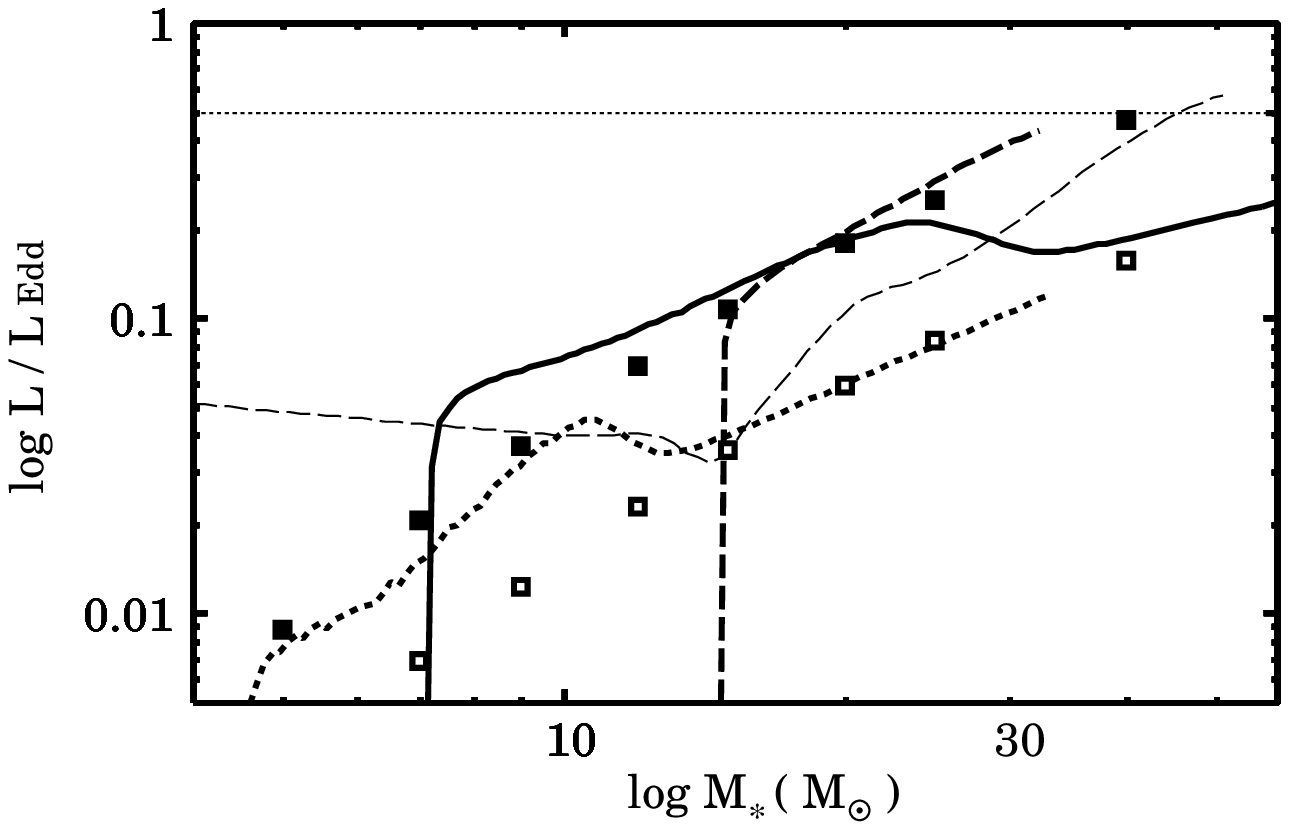}
\caption{Evolution of Eddington ratio of protostars
with disk accretion. The dotted, solid, and dashed lines represent
$L_\ast / L_{\rm Edd}$ for the different accretion rates
$\dot{M}_\ast = 10^{-4}~M_{\odot}~{\rm yr}^{-1}$ (case MD4-SDm0.1),
$10^{-3}~M_{\odot}~{\rm yr}^{-1}$ 
(case MD3-SDm0.1), and $4 \times 10^{-3}~M_{\odot}~{\rm yr}^{-1}$ 
(case MD4x3-SDm0.5), respectively.
The thin dashed line shows the evolution of $L_{\rm tot} / L_{\rm Edd}$
for spherical accretion with $4 \times 10^{-3}~M_{\odot}~{\rm yr}^{-1}$
(case MD4x3-S) for comparison, whereby 
$L_{\rm tot} = L_\ast + G M_\ast \dot{M}_\ast / R_\ast$.
The thin dotted line indicates the critical
ratio $L /L_{\rm Edd} = 0.5$. The open squares show the luminosity
of ZAMS stars $L_{\rm ZAMS} (M_\ast)$ calculated by Schaller et al. (1992).
The filled squares show $3 \times L_{\rm ZAMS} (M_\ast)$.}
\label{fig:ledd}
\end{figure*}

Figure \ref{fig:m_r_mdotv} shows a qualitatively different
evolution at the higher rate 
$\dot{M}_\ast = 4 \times 10^{-3}~M_{\odot}~{\rm yr}^{-1}$.
With spherical accretion a period of rapid expansion occurs when
$M_\ast \simeq 50~M_\odot$.
This expansion was originally found by Omukai \& Palla (2001, 2003)
at zero metallicity. 
The expansion occurs when the total luminosity $L_{\rm tot}$ 
approaches the Eddington luminosity, where 
$L_{\rm tot} = L_\ast + L_{\rm acc}$ and $L_{\rm acc}$ is 
an additional component from the accretion
shock front, $L_{\rm acc} = G M_\ast \dot{M}_\ast / R_\ast$.
In Paper I, we showed that the critical Eddington ratio 
$L_{\rm tot}/L_{\rm Edd}$, above which the expansion occurs, 
is about 0.5 at $Z=Z_\odot$ (presented again in Figure \ref{fig:ledd}).
The critical ratio less than unity is because the opacity near 
the stellar surface is somewhat higher than the electron-scattering 
opacity.

When $L_{\rm tot}$ reaches $0.5 L_{\rm Edd}$ in the KH contraction
phase, rapid expansion occurs and inhibits 
growth of the protostar by steady mass accretion.
We can derive the critical accretion rate $\dot{M}_{\rm \ast, cr}$,
with which the protostar barely reaches the ZAMS without undergoing
rapid expansion, from the critical Eddington ratio as follows
(Omukai \& Palla 2003; Paper I).
In the critical case, $L_{\rm tot}$ reaches $0.5 L_{\rm Edd}$
when the protostar arrives to the ZAMS. 
Thus, in the spherical accretion case, the critical rate 
$\dot{M}_{\rm \ast,cr}$ satisfies the relation
\begin{equation}
L_{\rm ZAMS} + \frac{G M_{\rm ZAMS} \dot{M}_{\rm \ast,cr}}{R_{\rm ZAMS}}
= 0.5 L_{\rm Edd} , 
\end{equation}
where the suffix ``ZAMS'' indicates the quantities of ZAMS stars.
This relation is transformed to
\begin{equation}
\dot{M}_{\rm \ast,cr} = \frac{4 \pi c R_{\rm ZAMS}}{\kappa_{\rm esc}}
              \left( 1 - \frac{L_{\rm ZAMS}}{0.5 L_{\rm Edd}} \right) ,
\end{equation}
where $\kappa_{\rm esc}$ is the electron-scattering opacity,
a lower limit to the actual opacity.
The critical rate given by this equation is a function of $M_\ast$,
but its dependence is weak.
The numerical value is $\dot{M}_{\rm \ast,cr} \simeq 
3 \times 10^{-3}~M_\odot~{\rm yr}^{-1}$ over a wide range of 
$M_\ast$, which is consistent with our numerical results.

We see that a similar expansion also occurs for
$4 \times 10^{-3}~M_{\odot}~{\rm yr}^{-1}$ in the cold disk 
accretion limit in Figure \ref{fig:m_r_mdotv}.
In this case the protostar very abruptly inflates when
$M_\ast \gtrsim 20~M_\odot$ and never turns to KH contraction. 
The stellar radius finally exceeds $600~R_\odot$ 
for $M_\ast \simeq 30~M_\odot$. YB08 also reported a similar 
evolution for the accretion rate $10^{-2}~M_{\odot}~{\rm yr}^{-1}$. 
Interestingly, it is known that a similar expansion
also occurs when a main-sequence star undergoes rapid mass accretion
due to mass exchange in close binary systems. 
Kippenhahn \& Mayer-Hofmeister (1977) and Neo et al. (1977) 
considered the accretion of deuterium-free gas onto existing 
main sequence stars.
The numerical method they adopted is basically the same as ours;
the protostellar structure was numerically calculated with a freely
radiating photospheric boundary condition.

Figure \ref{fig:ledd} shows that the critical Eddington ratio for 
the expansion is also $L_\ast / L_{\rm Edd} \simeq 0.5$ in the cold
disk accretion limit. 
Note that there is no additional luminosity component $L_{\rm acc}$ here. 
The accretion luminosity is assumed to have already radiated 
into free space before the accreted material settles onto the star.
The critical accretion rate in this case is derived as follows.
As explained in Section \ref{ssec:d_acc},  the stellar luminosity $L_\ast$ 
significantly rises in the swelling phase. 
The luminosity comes from release of gravitational energy after that. 
Figure \ref{fig:ledd} shows that the stellar luminosity in this phase
roughly follows
\begin{equation} 
L_\ast \simeq 3 \times L_{\rm ZAMS} (M_\ast) 
\label{eq:lst}
\end{equation}
in all cases considered.
The stellar luminosity approaches $L_{\rm ZAMS} (M_\ast)$ just before 
the arrival to the ZAMS at the rate $10^{-4}$ and 
$10^{-3}~M_\odot~{\rm yr}^{-1}$.
At the rate $4 \times 10^{-3}~M_\odot~{\rm yr}^{-1}$, however, 
$L_\ast$ approaches $0.5 L_{\rm Edd}$ before reaching the ZAMS
and the abrupt expansion occurs.
The stellar mass at which $L_\ast \simeq 0.5 L_{\rm Edd}$ is 
$M_\ast \simeq 45~M_\odot$ using equation (\ref{eq:lst}).
In order to avoid expansion the protostar has to reach
the ZAMS with a mass $M_{\rm \ast, ZAMS} \lesssim 45~M_\odot$.
This condition is transformed to that for accretion rates with
equation (\ref{eq:mzams}), 
\begin{equation}
\dot{M} \lesssim \dot{M}_{\rm \ast, cr} 
\simeq 2 \times 10^{-3}~M_\odot~{\rm yr}^{-1} .
\end{equation}
The derived critical rate is comparable to that for spherically symmetric
accretion, as confirmed by our numerical calculations. 

Our calculations suggest that a massive protostar significantly bloats
up and can not reach the ZAMS by steady mass accretion,
if the accretion rate is higher than a few 
$10^{-3}~M_\odot~{\rm yr}^{-1}$.
This feature is independent of the accretion geometry.
If mass accretion completely shuts off as a result of the 
fast expansion of the protostar, there will be an upper mass 
limit of pre-main-sequence stars around several 
10s$~M_\odot$, as discussed in Paper I.
Otherwise, mass accretion might continue in a non-steady fashion.
At least, it is certain that the radius of a massive protostar
reaches several 100s$~R_\odot$ at the high accretion rate.
The very large radius leads to a low effective temperature 
of the protostar.  

For case MD4x3-SDm0.5, for example, the stellar luminosity 
exceeds $10^5~L_\odot$ for $M_\ast \gtrsim 18~M_\odot$, but 
the effective temperature is only $T_{\rm eff} \lesssim 10^4$~K,
the value of a $M_\ast \lesssim 2.5~M_\odot$ ZAMS star. 
The stellar UV luminosity is very low for such low $T_{\rm eff}$.
This is important for the growth of an H~II region around
the protostar (e.g., Hoare \& Franco 2007).
Observationally, High-Mass Protostellar Objects (HMPOs) are 
considered to be young forming massive stars prior to the formation 
of an H~II region.        
However, a lot of HMPOs have high infrared luminosity exceeding
$10^4~L_\odot$ without observable H~II regions
(e.g., Sridharan et al. 2002; Beltr\'an et al. 2006b; 
Beuther et al. 2007). 
There has been some speculation regarding 
what hinders the growth of H~II regions 
around these bright sources.
ZAMS stars with such high luminosity should ionize their surroundings.
Some authors have proposed that rapid mass accretion quenches
the growth of H~II regions by increasing the gas density and 
recombination rate near the protostar (e.g., Walmsley 1995). 
With disk accretion, however, the H~II region would breaks out in the 
polar direction, where the gas density decreases significantly after the
material with low angular momentum has accreted onto the star or
been expelled in an outflow.

Our calculations well explain the existence of such 
bright HMPOs even with disk accretion. 
Massive stars accreting at high rates bloat up;
their UV luminosity is too low to form the H~II region.

\section{Summary and Conclusions}
\label{sec:sum} 

In this paper we have studied the evolution of massive protostars
with disk accretion at high rates 
$\dot{M}_\ast > 10^{-4}~M_{\odot}~{\rm yr}^{-1}$. 
We considered the limiting case of ``cold'' disk accretion, whereby
most of the stellar surface is not affected by the accretion flow,
and the accreting material brings a minimum of entropy
into the star. 
We calculated the evolution of protostars in this limiting case 
by adopting the photospheric (freely radiating) boundary condition.
The calculated evolution was compared and contrasted 
to the evolution with spherically symmetric accretion, 
corresponding to the opposite limit, whereby the accreting 
gas transports the maximum amount of entropy into the star. 

First, we considered in detail the evolution for
$\dot{M}_\ast = 10^{-3}~M_{\odot}~{\rm yr}^{-1}$.
The basic evolution via the cold disk accretion is summarized 
as follows:

The entire evolution is divided into four evolutionary phases:
(I) convection, (II) swelling, (III) KH contraction, and (IV) the
main-sequence accretion phase. 
The evolution in the first three phases varies with different
initial models. Roughly speaking, the stellar radius reaches 
its maximum $30 - 400~R_\odot$ for 
$M_\ast \simeq 10~M_\odot$ at the end of the swelling phase (II). 
On the other hand, the protostar begins the final phase (IV) 
for $M_\ast \simeq 30~M_\odot$, independent of different
initial models.

A key physical quantity governing the evolution is the
opacity in the stellar interior, analogous to the spherical 
accretion case. The opacity decreases with increasing
stellar mass, which causes the variety of the evolution.
In phase (I) the opacity is so high that the radiative heat
transport within the star does not affect the stellar structure.
In phase (II) the entropy accumulated in the stellar interior 
is gradually transported outward as the opacity decreases. 
A part of the transported entropy is received in a thin layer
near the stellar surface. Due to the high entropy deposited 
into this layer it significantly bloats up, resulting in the swelling
seen in phase (II).
Afterwards, the opacity decreases sufficiently, allowing 
the star to lose heat by efficiently radiation. 
The protostar contracts and the interior densities and
temperatures rise. 
This is the KH contraction phase (III).
The protostar finally reaches the main-sequence accretion phase
(IV), just after the active hydrogen burning begins. 

A major difference between the results for spherical accretion 
and cold disk accretion is the role of deuterium burning.
In the limit of cold disk accretion the accreting gas brings
a minimal amount of entropy into the star. 
Since the stellar radius is smaller because of the lower average 
entropy in the stellar interior, the protostellar radius is 
initially small, $R_\ast \simeq$ a few $\times R_\odot$. 
However, deuterium burning begins earlier, which raises the 
entropy content within the star. This enhances the swelling.
Consequently, the maximum radius is as large
as that with spherical accretion. 

Next, we calculated the protostellar evolution
at zero metallicity in the cold disk accretion limit. 
The protostellar evolution varies with metallicities in the
same manner as with spherical accretion; at $Z = 0$, the protostar 
swells up slightly earlier, and reaches the 
ZAMS with a smaller stellar radius.

Finally, we examined the evolution with different
accretion rates.
Our calculations show some dependences on accretion
rates which were also found with spherical accretion.
First, the stellar mass at the arrival to the ZAMS 
is higher for higher accretion rates. 
However, if the accretion rate is higher than a few 
$10^{-3}~M_\odot~{\rm yr}^{-1}$, the stellar radius
continues to increase without returning to KH contraction. 
The protostar never reaches the ZAMS by steady mass accretion
at such high rates.

The fact that a massive protostar bloats up to a radius of
several 100s$~R_\odot$ with an accretion rate exceeding
$10^{-3}~M_\odot~{\rm yr}^{-1}$ is a very robust result of
our studies.
The large radius leads to a low effective temperature and a 
correspondingly low stellar UV luminosity. 
Thus, the growth of an H~II region will be delayed until 
mass accretion onto the star deceases significantly. 
This explains the existence of HMPOs with high luminosities
exceeding $10^4~L_\odot$.

{\acknowledgements 
We thank Peter Bodenheimer for a careful critique of the manuscript.
We also thank Neal Turner for fruitful comments 
and discussions. 
This study is supported in part by Research Fellowships of the Japan
Society for the Promotion of Science for Young Scientists (TH) and 
by the Grants-in-Aid by the Ministry of Education, Science and 
Culture of Japan (18740117, 18026008, 19047004: KO).  
Portions of this work were conducted at the Jet Propulsion Laboratory,
California Institute of Technology, operating under a contract with 
the National Aeronautics and Space Administration (NASA).
}

\bibliographystyle{apj}
\bibliography{refs}

\clearpage


\appendix

\section{Evolution for $\dot{M}_\ast = 10^{-3}~M_{\odot}~{\rm yr}^{-1}$
with Shallow Initial Entropy Profile}
\label{ap:b0.1}

\begin{figure*}
\epsscale{0.7}
\plotone{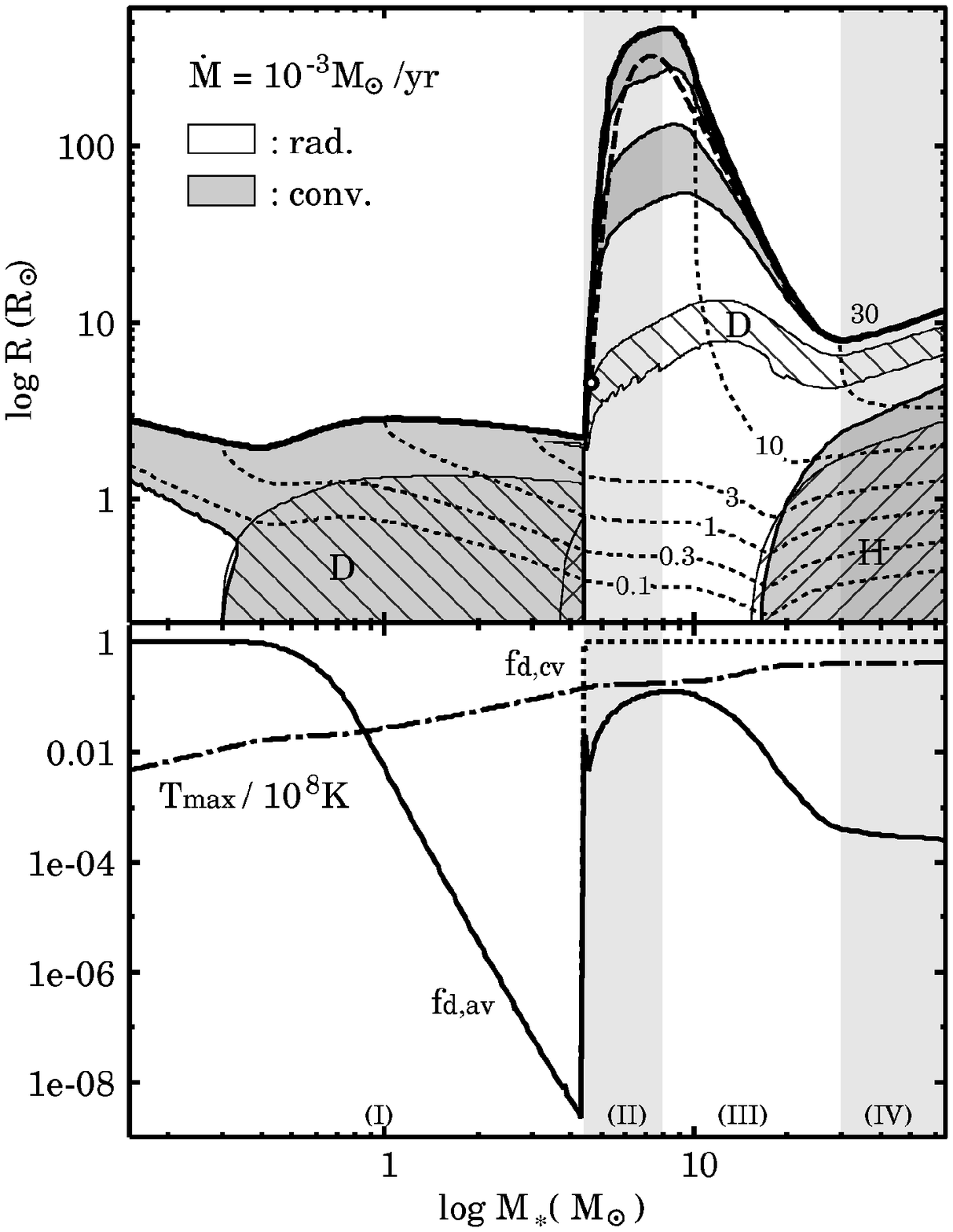}
\caption{
Same as Fig.\ref{fig:struct_log_1em3b1} but for a different initial
model with a shallower entropy distribution (case MD3-D-b0.1).
In the upper panel the dashed line shows the evolution of 
stellar radius after deuterium burning is turned off at the 
point marked by the open circle.}
\label{fig:str_log_1em3b0.1}
\end{figure*}

In Section \ref{sssec:imod}, we have seen that the protostellar
evolution differs with different initial models under the
cold disk accretion even at the same accretion rate. 
In this appendix, we add a more detailed explanation of these differences.
We revisit case MD3-D-b0.1 (Section \ref{sssec:imod}), 
whereby the initial entropy distribution is shallower than for the
fiducial case MD3-D.
The principle difference occurs in the first convection phase (I) 
(see Figure \ref{fig:str_log_1em3b0.1}). 
A central convective core emerges soon after the onset 
of deuterium burning when $M_\ast \simeq 0.3~M_\odot$. 
The convective core grows outward 
and merges with the outer convective layer near the surface:
The protostar becomes fully convective 
for $M_\ast \simeq 0.4~M_\odot$.
This is not seen in case MD3-D, where the 
core remains radiative even after D-burning begins. 

The subsequent evolution also differs.
For case MD3-D (see Figure \ref{fig:struct_log_1em3b1})
the radiative core gradually extends outward with increasing 
the stellar mass.
The D-burning layer, located at
the bottom of the outer convective layer, also moves outward,
until the protostar swells up ($M_\ast \simeq 6~M_\odot$).
For case MD3-D-b0.1, on the other hand, deuterium burning 
continues at the stellar center as long as the protostar 
remains fully convective and freshly accreted
gas (containing deuterium) is mixed downward. 
The deuterium is significantly depleted in this fully
convective phase (lower panel of Fig. \ref{fig:str_log_1em3b0.1}).
When $M_\ast \simeq 5~M_\odot$ a thin radiative layer 
(a so-called ``radiative barrier'') forms within the star, and
the stellar structure changes significantly.
This radiative barrier blocks the convective transport of newly accreted 
deuterium into the stellar interior. 
Without the refurbishment of deuterium, the central zones quickly
consume all available deuterium and become radiative.
Deuterium burning continues in a thin layer outside 
the radiative core.

The evolution described above is simlar to that of intermediate-mass
protostars as calculated by Palla \& Stahler 
(1991, 1992; also see Appendix \ref{ap:cp_ps} below).
The significant heat input by the deuterium shell-burning
triggers the rapid swelling of the star. 
The stellar radius ultimately expands to about $400~R_\odot$.
As in case MD3-D, however,  deuterium shell-burning per se
is not the main driver of the swelling. 
Figure \ref{fig:str_log_1em3b0.1} shows that, even if we turn off 
D-burning just after the swelling begins, the protostar continues 
to swell up.
The protostar expands because a thin surface layer bloats up
after receiving entropy transported from the inner part of the star.

\section{Comparison with Previous Work}
\label{ap:prev}

\subsection{Comparison with Palla \& Stahler (1992)}
\label{ap:cp_ps}

\begin{figure*}
\epsscale{0.7}
\plotone{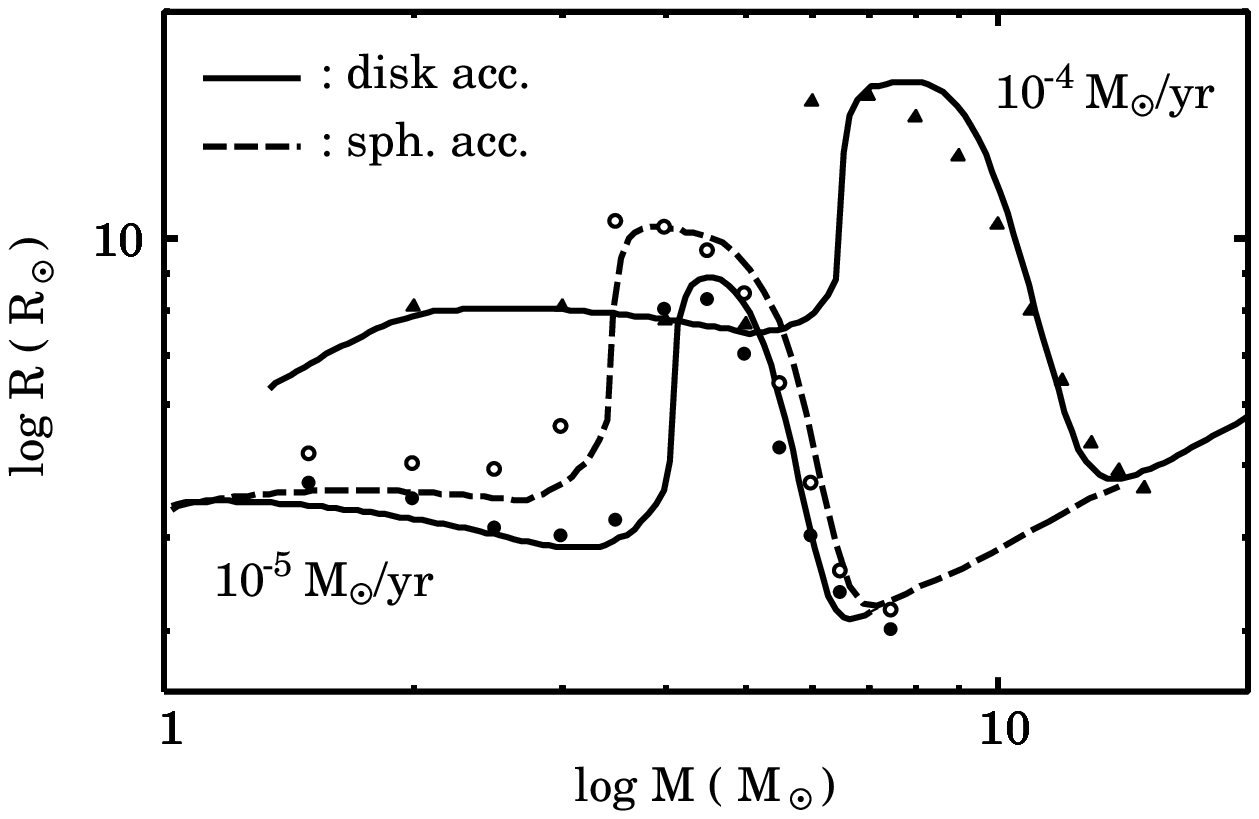}
\caption{Comparison with Palla \& Stahler (1992):
The evolution of protostellar radius for our calculations (lines) 
and Palla \& Stahler's (symbols). 
The dashed line and open circles represent the evolution 
via spherical accretion for $10^{-5}~M_{\odot} {\rm yr}^{-1}$ 
(case MD5-S-cv).
The evolution via disk accretion for 
$10^{-5}$ \& $10^{-4}~M_{\odot} {\rm yr}^{-1}$ 
(cases MD5 \& MD4-D-cv)
is depicted by the lower \& upper solid line and 
the filled circles \& triangles, respectively.
The initial models were constructed
following Palla \& Stahler (1992).}
\label{fig:m_r_ps}
\end{figure*}

We compare our numerical results with previous
calculations by Palla \& Stahler (1992), who studied 
protostellar evolution in the limit of cold disk accretion 
with accretion rates $10^{-5}~M_{\odot}~{\rm yr}^{-1}$ and
$10^{-4}~M_{\odot}~{\rm yr}^{-1}$.
The numerical method adopted by Palla \& Stahler (1992) is the
same as ours (e.g., Stahler, Shu \& Taam 1980). 
The same photospheric boundary condition 
(equations \ref{eq:psb_p} and \ref{eq:psb_l}) 
was used for the limiting case of cold disk accretion.  
The initial models they adopted are $1.0~M_\odot$ and $1.3~M_\odot$
fully convective stars at accretion rates of $10^{-5}~M_{\odot}~{\rm yr}^{-1}$
and $10^{-4}~M_{\odot}~{\rm yr}^{-1}$, respectively.
We also constructed the same initial models following their
method and calculated the subsequent evolution at each accretion
rate. Figure \ref{fig:m_r_ps} shows the resulting mass-radius 
relations. Our calculations agree well with the results presented by
Palla \& Stahler (1992).
For these cases the protostar remains fully convective 
until just before the swelling, when a radiative barrier
appears as for case MD3-D-b0.1 discussed above.

In section \ref{sec:mdot_dep} we discussed another case with
$10^{-4}~M_{\odot}~{\rm yr}^{-1}$ (case MD4-SDm0.1) in addition to
that shown in Figure \ref{fig:m_r_ps}.  Because the initial models differ,
the mass-radius relations are slightly different between these cases,
although the adopted accretion rate and boundary condition are the same.

\subsection{Comparison with Yorke \& Bodenheimer (2008)}
\label{ap:cp_yb}

\begin{figure*}
\epsscale{0.7}
\plotone{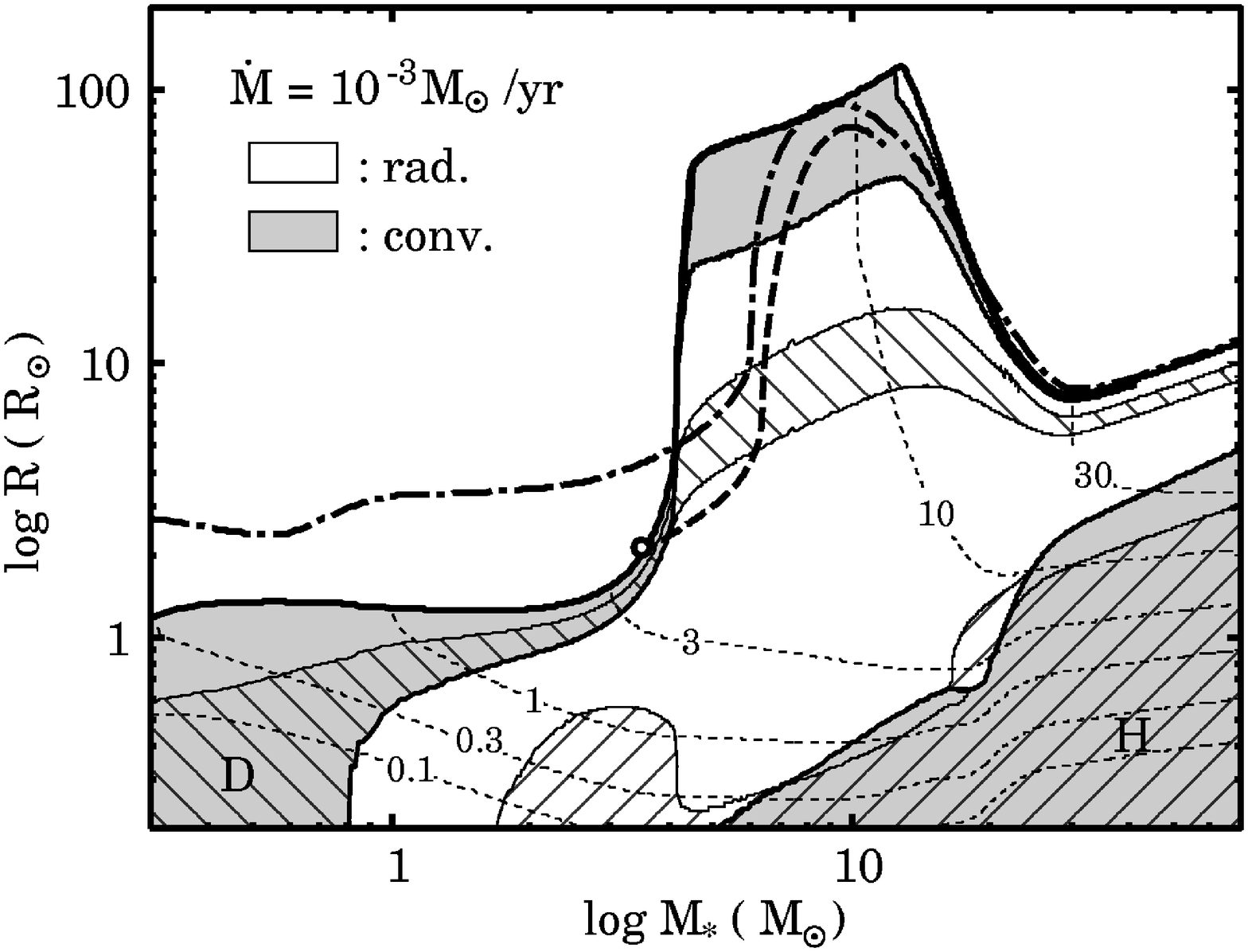}
\caption{Comparison with Yorke \& Bodenheimer (2008):
The evolution of protostellar radius and 
interior structure calculated by Yorke \& Bodenheimer (2008)
for an adopted accretion rate
$\dot{M}_\ast = 10^{-3}~M_{\odot}~{\rm yr}^{-1}$.
The interior structure is illustrated in the same manner
as in the upper panel of Figure \ref{fig:str_log_1em3b1_sh}. 
The dashed line represents the evolution of the radius
after turning off all nuclear reactions when $M_\ast \simeq 3.5~M_\odot$,
indicated on the plot by the open circle.
The dot-dashed line shows the evolution of the  
radius in case MD3-D (taken from Fig. \ref{fig:struct_log_1em3b1}).}
\label{fig:struct_log_yb}
\end{figure*}

Next, we compare the results presented in Section \ref{ssec:d_acc} 
to the previous calculations by YB08 (see also discussion in 
Section \ref{sec:intro}).  YB08 independently calculated
the evolution of accreting massive protostars by solving
the interior structure using a different numerical code than
that used by Palla \& Stahler (1992).
To make the comparison clear we recalculate the evolution 
with their code and briefly analyze the
evolution of the stellar interior structure. 
The basic equations are the usual four stellar structure 
equations, taking into account the effect of mass accretion.
The adopted boundary condition slightly differs from the one
depicted in equations (\ref{eq:psb_p}) and (\ref{eq:psb_l}), but 
is essentially the same photospheric boundary condition 
(Bodenheimer et al. 2007; YB08).  YB08 solve for a spherical 
grey stellar atmosphere including curvature effects. 
The stellar interior models are constructed using the Henyey method
iterated to match the grey atmosphere.
The initial model is a fully convective $0.1~M_\odot$ star of radius
$1~R_\odot$.

Figure \ref{fig:struct_log_yb} shows the evolution of the 
stellar radius and interior structure 
for $\dot{M}_\ast = 10^{-3}~M_{\odot}~{\rm yr}^{-1}$.
The protostar is initially fully convective, until a
radiative core appears for $M_\ast \simeq 0.8~M_\odot$.
The subsequent evolution is very similar to that for case MD3-D
discussed in Section \ref{sssec:fid} 
(see Fig. \ref{fig:struct_log_1em3b1} for comparison). 
The radiative core gradually grows outward as the stellar mass
increases. The D-burning layer, located at the
bottom of the outer convective layer, also moves outward. 
The stellar radius remains $\sim 1~R_\odot$ during this phase,
which is slightly smaller than for case MD3-D.  At this 
smaller radius (and higher central densities and temperatures)
central hydrogen burning occurs earlier than for case MD3-D.
The protostar rapidly swells up when $M_\ast \simeq 4~M_\odot$.
The radiative core simultaneously expands to cover
most of the star.  Central hydrogen burning nearly dies out.

As noted in Section \ref{sssec:fid}, deuterium burning itself 
is not the cause of the swelling.
Figure \ref{fig:struct_log_yb} shows that even if all nuclear 
reactions are turned off when $M_\ast \simeq 3.5~M_\odot$, the 
protostar significantly swells up later.
After the stellar radius reaches its maximum of $120~R_\odot$ when
$M_\ast \simeq 13~M_\odot$, the protostar contracts and
hydrogen burning increases in the center.
The convective core also grows due to the increased 
heat input by hydrogen burning.
The star reaches the ZAMS for $M_\ast \simeq 30~M_\odot$
and continues to closely follow the ZAMS as it accretes new material.
All these evolutionary features are also seen in case MD3-D. 
We conclude that the evolution presented in Section \ref{ssec:d_acc} 
is fully consistent with the previous calculation by YB08.

\end{document}